\documentclass[prc,twocolumn,preprintnumbers,showpacs]{revtex4}
\usepackage{gensymb}
\usepackage{amsmath,amssymb,epsfig,bm,mathrsfs,color}
\usepackage{enumerate}
\usepackage[colorlinks=true,linkcolor=blue,citecolor=blue,urlcolor=blue]{hyperref}
\usepackage{slashed}
\usepackage{color}
\newcommand{\be}{\begin{equation}}
\newcommand{\ee}{\end{equation}}
\newcommand{\bea}{\begin{eqnarray}}
\newcommand{\eea}{\end{eqnarray}}

\newcommand{\veck}{{\bm k}}
\newcommand{\vecQ}{{\bm Q}}

\newcommand{\vecr}{\bm r}

\newcommand{\SD}{^3S_1$-$^3D_1}

\definecolor{red}{rgb}{0.8,0,0}
\definecolor{orange}{rgb}{0.8,0.2,0.0}
\definecolor{blue}{rgb}{0.3,0.0,0.8}
\definecolor{green}{rgb}{0,0.5,0.0}
\definecolor{darkred}{rgb}{0.7,.1,.2}
\definecolor{bgred}{rgb}{1.,.95,.95}
\definecolor{bgblue}{rgb}{.95,.95,1.}
\definecolor{bluegreen}{rgb}{0.,.5,.3}
\definecolor{darkred}{rgb}{0.7,.1,.2}
\definecolor{darkgreen}{rgb}{0.1,.6,.0}
\definecolor{lightyellow}{rgb}{1.,1.,.8}
\definecolor{darkcyan}{rgb}{0.,.7,.9}
\definecolor{lightblue}{rgb}{0.6,0.8,1}
\definecolor{lightgreen}{rgb}{0.7,1.,.9}
\definecolor{money}{rgb}{0.4,0.8,0.}
\definecolor{purple}{rgb}{0.9,0.0,0.8}
\definecolor{orange}{rgb}{0.9,0.5,0.0}
\definecolor{newgr}{rgb}{0.2,0.8,0.2}
\definecolor{newbl}{rgb}{0.3,0.6,0.8}
\definecolor{newor}{rgb}{1.0,0.6,0.}

\def\mnras{MNRAS}%
\def\prb{Phys.~Rev.~B}%
\def\prc{Phys.~Rev.~C}%
\begin{document}
\title{
Spin-polarized neutron matter: Critical unpairing  and BCS-BEC precursor
}

\author{Martin Stein}
\email{mstein@th.physik.uni-frankfurt.de}
\affiliation{
{
Institute for Theoretical Physics,
  J.~W.~Goethe-University, D-60438  Frankfurt am Main, Germany
}}

\author{Armen Sedrakian}
\email{sedrakian@th.physik.uni-frankfurt.de}
\affiliation{
{
Institute for Theoretical Physics,
J.~W.~Goethe-University, D-60438  Frankfurt am Main, Germany
}}

\author{Xu-Guang Huang}
\email{huangxuguang@fudan.edu.cn}
\affiliation{
{
Physics Department \& Center for Particle Physics and
  Field Theory, \\Fudan University, Shanghai 200433, China
}}
\author{John W.~Clark}
\email{jwc@wuphys.wustl.edu}
\affiliation{
{
Department of Physics and McDonnell Center for the Space 
Sciences, Washington University, St.~Louis, Missouri 63130, USA
}}
\affiliation{
{
Centro de Ci\^encias Matem\'aticas, 
University of Madeira, 9000-390 Funchal, Portugal
}}

\date{\today}

\begin{abstract}
  We obtain the critical magnetic field required for complete
  destruction of $S$-wave pairing in neutron matter, thereby setting
  limits on the pairing and superfluidity of neutrons in the crust and
  outer core of magnetars. We find that for fields $B \ge 10^{17}$ G
  the neutron fluid is non-superfluid -- if weaker spin-1 superfluidity
  does not intervene -- a result with profound consequences for the
  thermal, rotational, and oscillatory behavior of magnetars.  Because
  the dineutron is not bound in vacuum, cold dilute neutron matter
  cannot exhibit a proper BCS-BEC crossover.  Nevertheless, owing to
  the strongly resonant behavior of the $nn$ interaction at low
  densities, neutron matter shows a precursor of the BEC state, as
  manifested in Cooper-pair correlation lengths {being} comparable to
  the interparticle distance.  { We make a systematic quantitative
    study of this type of BCS-BEC crossover in the presence of neutron
    fluid spin-polarization induced by an ultra-strong magnetic
    field. We evaluate the Cooper pair wave-function, quasiparticle
    occupation numbers, and quasiparticle spectra for densities and
    temperatures spanning the BCS-BEC crossover region.  The phase
    diagram of spin-polarized neutron matter is constructed and
    explored at different polarizations.}
\end{abstract}

\maketitle

\section{Introduction}

A comprehensive understanding of the thermodynamic properties of
strongly magnetized baryonic matter is one of the major challenges in
the astrophysics of compact stars. There is substantial observational
evidence that anomalous x-ray pulsars and soft-$\gamma$-ray repeaters are
two manifestations of strongly magnetized neutron stars, known as
magnetars, which are characterized by surface fields of order
$B\sim 10^{15}$ G~\cite{1995MNRAS.275..255T}.
These identifications are consistent with the measured slow spins and
large spin-down rates as well as with the energetics of observed
magnetic activity associated with flares.  Magnetic fields play a
secondary role in the structure and thermal emission of ordinary
neutron stars with fields $B\sim 10^{12}$~G, whereas the fields in
magnetars are large enough to impact basic physical properties of the
stellar matter, including its equation of state, its crust composition,
and its pairing and superfluid properties.

In  {this work we focus} on the behavior of pure 
neutron matter in strong magnetic fields. Specifically, we have carried 
out a detailed study of $S$-wave pairing in neutron matter as it 
exits at relatively low densities. Our results are of two-fold 
interest.  Firstly, we compute the critical magnetic field for unpairing 
of the $S$-wave condensate owing to the spin-alignment induced by 
the strong magnetic field, as measured by the spin polarization. 
The results are of direct practical interest for the astrophysics 
of magnetars, as the derived critical field for pair disruption
limits the occurence of neutron superfluidity in the low-density
(outer core and crust) regions of a neutron star. Second, we study
the signatures of a BCS-BEC crossover~\cite{1985JLTP...59..195N} in
dilute neutron matter and the emergence of dineutron correlations in
a magnetic field, thus generalizing to the case of spin-polarized
neutron matter the previous studies of this clustering 
phenomenon in infinite neutron
matter~\cite{2006PhRvC..73d4309M,2007PhRvC..76f4316M,2009PhRvC..79e4003A,2013NuPhA.909....8S,2014PAN....77.1145K}
and in finite nuclear systems~\cite{2009PhRvC..79e4305K}.

The BCS-BEC crossover, in the sense of Nozi{\`e}res-Schmitt-Rink
theory~\cite{1985JLTP...59..195N}, occurs naturally in the $\SD$
channel in isospin-symmetric~\cite{1993NuPhA.551...45A,1995PhRvC..52..975B,2006PhRvC..73c5803S,2010PhRvC..81c4007H,2012PhRvC..86a4305S}
and isospin-asymmetric~\cite{2001PhRvC..64f4314L,2012PhRvC..86f2801S,2014PhRvC..90f5804S} nuclear matter,
where the bound pairs are deuterons in the low-density limit.
Neutron-neutron ($nn$) pairing in the $^1S_0$ channel comes into play in
nuclear matter when the isospin asymmetry of the system is large
enough to suppress the (otherwise dominant) attractive interaction the
$\SD$ pairing channel.  In pure neutron matter, isospin-triplet
pairing in the $\SD$ channel is prohibited by Pauli blocking; hence
the dominant pairing channel must be an isospin-singlet state,
necessarily $^1S_0$ in the low-density regime, as implied by the 
nuclear phase-shift analysis (see~Ref.~\cite{2006pfsb.book..135S}).

The primary effect of a magnetic field on a neutron Cooper pair
is the alignment of their spins caused by the Pauli paramagnetic 
interaction between $B$ field and the spin magnetic moments of 
the neutrons.  Plainly, a large enough magnetic field will
quench pairing.  This $nn$ pair-breaking effect may be contrasted
with that for proton pairs (and ultimately charged hyperons), 
which become unpaired at lower field strengths owing to Landau
diamagnetic currents~\cite{AM,2015PhRvC..91c5805S,2014arXiv1403.2829S}. 

The present description is constrained to the low-density regime below
the saturation density of symmetrical nuclear matter, $\rho_0 = 0.16$
fm$^{-3}$.  At higher densities the dominant pairing state in neutron
matter shifts to the $^3P_2$-$^3F_2$ channel, which induces a spin 1
condensate of neutrons~\cite{2006pfsb.book..135S}. In this case, the
spin-polarizing effect of the magnetic field on the internal structure
of the spin-1 pairs is nondestructive.

The two-neutron system has no bound state in vacuum, so dilution of
neutron matter does not lead automatically to a state populated by
tightly bound dineutrons that could undergo Bose-Einstein condensation
(BEC).  { (It should be noted, however, that the bare neutron-neutron
  interaction in the $^1S_0$ channel supports a virtual state close to
  zero energy, which is characterized by a large scattering length
  $-18.5 \pm 0.4$ fm. It thus implies a strongly correlated
  $^1S_0$-wave state at asymptotically low densities).}  Nevertheless,
on general grounds one cannot expect a Bose condensate regime of
neutrons to be present in the low-density limit. This situation stands
in contrast to that for $\SD$ neutron-proton ($np$) pairing, where the
phase diagram exhibits both a BCS-BEC crossover region and a
well-defined Bose condensate of deuterons at asymptotically low
density. Notwithstanding the arguments above, it was shown in
Refs.~\cite{2006PhRvC..73d4309M,2007PhRvC..76f4316M,2009PhRvC..79e4003A,2013NuPhA.909....8S}
that a BCS-BEC {\it crossover region} may also arise in neutron matter
under dilution.  In principle, this phenomenon occurs in full analogy
to its counterpart for $\SD$ pairing, with the exception that the
asymptotical state of the system at low densities is a weakly
interacting neutron gas, instead of a Bose condensate of neutron
dimers.

The phase diagram of dilute neutron matter may contain anisotropic or
non-homogeneous phases such as the Larkin-Ovchinnikov\--Fulde-Ferrell
(hereafter LOFF) phase or a phase-separated phase~ (see
Refs.~\cite{2012PhRvC..86f2801S,2014PhRvC..90f5804S} and references
therein).  Below we provide a theoretical framework which incorporates
such phases; however, our numerical studies are confined to
homogeneous, isotropic solutions.

Neutron-neutron pairing plays a prominent role in the physics of the
inner crust of a neutron star~(Ref.~\cite{2006pfsb.book..135S} and
references therein).  Other systems characterized by strong neutron
excess are neutron-rich nuclei near the drip
line~\cite{1991AnPhy.209..327B,2013PhRvC..88c4314P,2014PhRvC..90c4313Z} 
and halo nuclei such as
$^{11}$Li~\cite{2010JPhG...37f4040H} that feature halo neutrons.  There
are conspicuous phenomenological signatures of neutron superfluidity
in neutron stars, providing strong evidence that a neutron pairing
condensate in the star's inner crust plays a prominent role in
neutrino cooling and in glitch-type timing behavior in
pulsars~\cite{2006pfsb.book..135S}.

This paper is structured as follows. In Sec.~\ref{sec:theory} we give
a brief presentation of the theory of spin-polarized neutron matter in
terms of imaginary-time finite-temperature Green's functions. In
Sec.~\ref{sec:phases} we discuss the results of extensive calculations
based on this microscopic many-body approach, namely the phase diagram
of neutron matter over the relevant low-density domain at various
degrees of polarization, the temperature-polarization dependence 
of the gap in the weak-coupling regime, the kernel of the gap 
equation in various coupling regimes, the Cooper-pair wave function 
across the BCS-BEC crossover, and quasiparticle occupation numbers and 
dispersion relations.  Section~\ref{sec:magentars} is concerned with 
the critical magnetic field required for unpairing of the condensate 
in the context of magnetars. Readers interested only in astrophysical 
implications of  {this} work can skip directly to this section. 
Our conclusions are summarized in Sec.~\ref{sec:conclude}.

\section{Theory}
\label{sec:theory}

The theory of spin-polarized pair-correlated neutron matter in
equilibrium can be formulated in the language of the imaginary-time
Nambu-Gorkov matrix Green's function
\bea 
\label{props} i\mathscr{G}_{12} =
i\left(\begin{array}{cc} G_{12}^{+} & F_{12}^{-}\\
    F_{12}^+ & G_{12}^{-}\end{array}\right) = \left(\begin{array}{cc}
    \langle T_\tau\psi_1\psi_2^+\rangle
    & \langle T_\tau\psi_1\psi_2\rangle \\
    \langle T_\tau\psi_1^+\psi_2^+\rangle & \langle T_\tau\psi_1^+\psi_2\rangle
\end{array}\right),
\eea
where the indices $1,2,\dots$ stand for the continuous space-time
variables $x = (t,\vecr) $ of the neutrons, thus 
$G_{12}^{+}\equiv G^{+}_{\alpha\beta}(x_1,x_2)$, etc., and greek
indices label discrete variables in general. In spin-polarized neutron
matter the isospin is fixed, so within the discrete nucleonic degrees 
of freedom only the Pauli spins play a dynamical role.  Therefore, each 
operator in Eq.~(\ref{props}) is a spinor, e.g.,
$\psi_{\alpha}=(\psi_{n\uparrow},\psi_{n\downarrow})^T$,
where the internal variables $\uparrow, \downarrow$ denote a 
particle's spin state. Accordingly, the propagators live in a $4\times 4$ 
space owing to the doubling of degrees of freedom in the Nambu-Gorkov 
formalism and owing to the breaking of the spin $SU(2)$ symmetry. 

The matrix {propagator} (\ref{props}) obeys the standard Dyson
equation, which we write in momentum space as
\be  
\label{eq:Dyson2} 
\left[\mathscr{G}_0(k,\vecQ)^{-1}-\Xi(k,\vecQ) \right]\mathscr{G}
(k,\vecQ) = {\bf{1}}_{4\times 4},
\ee  
where $\Xi(k,\vecQ) $ is the matrix self-energy.  To accommodate in
our formalism the  appearance of the LOFF phase, we do not assume 
translational invariance from the outset.  Hence the Green's 
functions and self-energies are allowed to depend on the 
center-of-mass momentum $\vecQ$ of Cooper pairs.  The relative 
(four-)momentum of pairs is of the form $k \equiv (ik_{\nu},\veck)$, 
in which the zeroth component assumes discrete values $k_{\nu} 
= (2\nu+1)\pi T $, where $\nu\in \mathbb{Z}$ and $T$ is the temperature.
Further reductions are possible by virtue of the fact that the normal
propagators for the particles and holes are diagonal in the spin
space, the off-diagonal elements of the free matrix propagator
$\mathscr{G}_0^{-1}$ being zero.  Writing out the non-vanishing
components in the Nambu-Gorkov space explicitly, we obtain
\begin{eqnarray}
\label{inverseG}
\mathscr{G}_0^{-1} &=&\begin{pmatrix}
   ik_{\nu} - \epsilon_{\uparrow}^+&0&0& 0\\
   0&ik_{\nu}- \epsilon_{\downarrow}^+&0 &0\\ 
   0&0 &ik_{\nu} + \epsilon_{\uparrow}^-&0\\ 
    0 &0  &0&ik_{\nu}+ \epsilon_{\downarrow}^-\\ 
 \end{pmatrix},\nonumber\\
\end{eqnarray}
where
\bea \epsilon^{\pm}_{\uparrow/\downarrow} &=&
\frac{1}{2m^*}\left(\veck\pm
  \frac{\vecQ}{2}\right)^2-\mu_{\uparrow/\downarrow}. 
\eea 
These single-particle energies can be separated into symmetrical and 
anti-symmetrical parts with respect to time-reversal operation by writing
\bea
\label{eq:spectra1}
\epsilon_{\uparrow}^{\pm} &=&  E_S-\delta\mu\pm E_A,\\
\label{eq:spectra2}
\epsilon_{\downarrow}^\pm &=& E_S+\delta\mu \pm E_A,
\eea 
where 
\bea
\label{eq:E_S}
E_S &=&\frac{Q^2/4+k^2}{2m^*}-\bar\mu, \\
\label{eq:E_A}
E_A &=& \frac{\veck\cdot \vecQ}{2m^*} ,
\eea 
are respectively the spin-symmetrical and spin-antisymmetrical parts of the 
quasiparticle spectrum and $\delta \mu \equiv (\mu_\uparrow - \mu_\downarrow)/2$ 
determines the shifts of chemical potentials of up-spin and down-spin 
neutrons from the mean $\bar\mu \equiv (\mu_\uparrow+\mu_\downarrow)/2$.  
The effective mass $m^*$ is  { computed from a Skyrme
  density functional, with SkIII \cite{1987PhRvC..35.1539S} and
SLy4 \cite{1998NuPhA.635..231C} parametrizations yielding nearly
identical results.
}
The quasiparticle spectra in Eq.~(\ref{inverseG}) are written in a 
general reference frame moving with the center-of-mass momentum $\vecQ$ 
relative to a laboratory frame at rest.  { The spectrum of quasiparticles 
is seen to be two-fold split owing  to finite $\vecQ$ and further split
owing to spin polarization, which breaks the spin 
$SU(2)$ internal symmetry of neutron matter.}

As already stressed in the Introduction, low-density neutron matter
interacts attractively in the $^1S_0$ channel, leading to isovector $nn$
spin-singlet pairing.  Accordingly, the anomalous propagators have the
property $(F^+_{12},F^-_{12})\propto i\sigma_y $, where $\sigma_y$ is
the second Pauli matrix in spin space. This implies that in the quasiparticle
approximation, the self-energy $\Xi$ has only off-diagonal elements in
Nambu-Gorkov space.  The inverse full Green's function on the 
left-hand-side of Eq.~(\ref{eq:Dyson2}) is then given by
\begin{widetext}
\begin{eqnarray}\mathscr{G}^{-1} = 
\mathscr{G}_0^{-1}-\Xi &=&\begin{pmatrix}
   ik_{\nu} - \epsilon_{\uparrow}^+&0&0&i\Delta\\
   0&ik_{\nu}- \epsilon_{\downarrow}^+&-i\Delta&0\\ 
   0&i\Delta &ik_{\nu} + \epsilon_{\uparrow}^-&0\\ 
   -i\Delta&0  &0&ik_{\nu}+ \epsilon_{\downarrow}^-\\ 
 \end{pmatrix}.
\end{eqnarray}
Thus, the Dyson equation takes the form
\begin{eqnarray}
\begin{pmatrix}
    ik_{\nu} - \epsilon_{\uparrow}^+&0&0&i\Delta\\
    0& ik_{\nu} - \epsilon_{\downarrow}^+ &-i\Delta&0\\ 
    0&i\Delta &ik_{\nu} + \epsilon_{\uparrow}^-&0\\ 
    -i\Delta&0&0&ik_{\nu} + \epsilon_{\downarrow}^-
  \end{pmatrix}\cdot
  \begin{pmatrix}
    \begin{matrix} G^+_\uparrow&0\\0&G^+_\downarrow \end{matrix} &
    \begin{matrix} 0&F_{\uparrow\downarrow}^-\\F_{\downarrow \uparrow}^-&0 \end{matrix} \\
    \begin{matrix} 0&F_{\uparrow\downarrow}^+\\F_{\downarrow \uparrow}^+&0 \end{matrix} &
    \begin{matrix} G^-_\uparrow&0\\0&G^-_\downarrow \end{matrix} &
  \end{pmatrix}=   {\rm diag}(1,1,1,1), 
\end{eqnarray}
\end{widetext}
 {where we use short-hand $G^+_\uparrow \equiv G^+_{\uparrow\uparrow}$} and so
on. 
The solutions of this equation provide the normal and anomalous
Green's functions
\bea
G_{\uparrow/\downarrow}^{\pm} &=&
\frac{ik_{\nu}\pm\epsilon_{\downarrow/\uparrow}^{\mp}}{(ik_{\nu}-E^+_{\mp/\pm})(ik_{\nu}+E^-_{\pm/\mp})},\\
F_{\uparrow\downarrow}^{\pm} &=&
\frac{-i\Delta}{(ik_{\nu}-E^+_{\pm})(ik_{\nu}+E^-_{\mp})},\\
F_{\downarrow\uparrow}^{\pm} &=&
\frac{i\Delta}{(ik_{\nu}-E^+_{\mp})(ik_{\nu}+E^-_{\pm})},
\eea
where the four branches of the quasiparticle spectrum are given by
\be
E_{r}^{a} = \sqrt{E_S^2+\Delta^2} + r\delta\mu +a E_A,
\ee
in which $a, r \in \{+,-\}$.  
In mean-field approximation, the
anomalous self-energy (pairing-gap) is determined by
\bea\label{eq:gap0}
\Delta({\veck},{\vecQ}) &=&  \frac{T}{4}\int\!\!\frac{d^3k'}{(2\pi)^3}
\sum_\nu V({\veck},{\veck}')\nonumber\\
&&\times {\rm Im} [F_{\uparrow\downarrow}^+(k'_\nu,{\veck}',{\vecQ})
+F_{\uparrow\downarrow}^-(k'_\nu,{\veck}',{\vecQ})\nonumber\\
&&-F_{\downarrow\uparrow}^+(k'_\nu,{\veck}',{
\vecQ})-F_{\downarrow\uparrow}^-(k'_\nu,{\veck}',{\vecQ})],\nonumber\\
\eea
where $ V(\veck,\veck')$ is the neutron-neutron interaction potential.
After partial-wave expansion in the potential we keep the $^1S_0$
component, compute the Matsubara sum and continue analytically to the
real axis; as a result we find the gap equation
\bea\label{eq:gap}
  \Delta(Q)&=&\frac{1}{4}\sum_{a,r}\int\frac{d^3k'}{(2\pi)^3} V(k,k')\nonumber\\
  &\times&\frac{\Delta(k',Q)}{2\sqrt{E_{S}^2(k')+\Delta^2(k',Q)}}[1-2f(E^a_r)],
\eea
where $f(E)$ is the Fermi function.
The densities of up-spin and down-spin particles are given by
\bea\label{eq:densities0}
  \rho_{\uparrow/\downarrow}({\vecQ})&=& T
  \int \frac{d^3k}{(2\pi)^3} \sum_\nu G_{\uparrow/\downarrow}^+(k_\nu,{\veck},{\vecQ}).
\end{eqnarray}
Performing the same operations as for the gap function, we obtain 
\bea\label{eq:densities}
 \rho_{\uparrow/\downarrow}(Q)
  &=&\int \frac{d^3k}{(2\pi)^3}\left[\frac{1}{2}\left(1+\frac{E_S}{\sqrt{E_S^2+\Delta^2}}\right)f(E^+_\mp)\right.\nonumber\\
&+&\left.\frac{1}{2}\left(1-\frac{E_S}{\sqrt{E_S^2+\Delta^2}}\right)(1-f(E^-_\pm))\right]\,.
\eea

At finite temperature $T$, the system minimizes its free energy 
by choosing the optimal values of the magnitude $Q$ of the 
center-of-mass momentum and the gap in Eqs.~(\ref{eq:gap}) and (\ref{eq:densities}). 
As a reference free energy we use the same quantity evaluated in 
the normal state with $Q$ and $\Delta$ both zero, labeling it
with an $N$ subscript as opposed to the $S$ subscript used for the
superfluid state.   Thus,
\be \label{eq:free}
{\cal F}_S =  {\cal E}_S-T{\cal S}_S
\quad {\rm versus} \quad 
{\cal F}_N =  {\cal E}_N-T{\cal S}_N,
\ee
where ${\cal E}$ denotes the internal energy (statistical average of 
the system Hamiltonian)  {and} ${\cal S}$ the entropy. 

We measure the spin polarization by the parameter
\be\label{eq:polarization} \alpha =
\frac{\rho_\uparrow-\rho_\downarrow}{\rho_\uparrow+\rho_\downarrow},
\ee 
where $\rho_\uparrow$ and $\rho_\downarrow$ are, respectively, the
number densities of the up-spin and down-spin components of the
neutron-matter system and $\rho = \rho_\uparrow + \rho_\downarrow$ is its
total particle (or baryon) density. The possible solutions, or
phases, of the variational problem so defined can be classified 
according to the alternatives
\bea\label{eq:phases}
\begin{array}{llll}
Q = 0,  &\Delta \neq 0, & x = 0,& \textrm{BCS phase,}\\
Q = 0, & \Delta = 0, & x = 1, &\textrm{unpaired phase,} \\
Q \neq 0, & \Delta \neq 0, &x = 0,&\textrm{LOFF phase,} \\
Q = 0, & \Delta\neq 0 , & 0< x < 1, &\textrm{phase-separated phase.} \\
\end{array}
\eea 
  The ground state corresponds to the phase with lowest free
  energy. Below we will consider exclusively homogeneous, isotropic solutions
  corresponding, respectively, to the first two lines in
  Eq. \eqref{eq:phases}; i.e.,
  the anisotropic or inhomogeneous phases (the LOFF phase and the phase-separated
  phase) is not considered.  

\section{BCS phase, search for LOFF phase, and crossover to BEC}
\label{sec:phases}
\begin{table*}[b,t]
\begin{tabular}{cccccccc}
\hline 
$\textrm{log}_{10}\left({\rho}/{\rho_0}\right)$ & $k_F$ [fm$^{-1}$] & $\Delta$ [MeV] & $m^\ast/m$ & $\mu_n$ [MeV] & $d$ [fm] & $\xi_{\rm rms}$ [fm] & $\xi_{a}$ [fm] \\
\hline\hline 
   $-1.0$ & 0.78 & 2.46 & 0.967 &12.94 & 2.46 & 4.87 & 4.33 \\
$-1.5$ & 0.53 & 1.91 & 0.989 & \,\,\,5.65 & 3.61 & 3.55 & 3.71 \\
$-2.0$ & 0.36 & 1.07 & 0.997 & \,\,\,2.49 & 5.30 & 2.36 & 4.48 \\
\hline\\
\end{tabular}
\caption{(Color online)
Tabulated values of characteristic parameters related to the $^1S_0$ 
condensate in dilute, unpolarized neutron matter at temperature 
$T=0.25$ MeV, for selected values of the total particle density 
$\rho$ (in units of the nuclear saturation density).  Other table
entries: Fermi momentum $k_F=(3\pi^2 \rho)^{1/3}$, pairing gap $\Delta$, 
effective mass (in units of bare mass), chemical potential $\mu_n$, 
interparticle distance $d$, and coherence lengths $\xi_{\rm rms}$ and 
$\xi_{a}$. 
}
\label{table_1}
\end{table*}
Based on calculations performed within the theoretical framework
summarized in Sec.~\ref{sec:theory}, we have generated the
temperature-density ($T-\rho$) phase diagram of dilute neutron matter
at various spin polarizations [Eq.~\eqref{eq:polarization}]. A number of key
quantities of the neutron condensate were studied at fixed $T$ and
$\rho$ corresponding to the different coupling strengths which
characterize the BCS versus quasi-BEC nature of the condensate.
Table~\ref{table_1} collects several quantities of interest at 
fixed $T=0.25$ MeV and vanishing spin polarization $\alpha$, 
for three values of the density $\rho$ that span the  {regimes studied
numerically. } The computations were carried out for the
rank 3 separable Paris potential (PEST~3) in the $^1S_0$ partial-wave
channel, with parameters given in Ref.~\cite{PhysRevC.30.1822}.  

Our findings concerning the BCS-BEC crossover are the following.  No
change of sign of the chemical potential was observed.  The chemical
potential $\bar\mu$ remains positive down to the lowest density
considered.  Specifically, the lowest value found for $\bar\mu$, 0.24
MeV, was obtained at the point $\ln (\rho/\rho_0) = -3.57$ and
$T = 0.05$ where $\Delta$ vanishes within the numerical accuracy of
our model. Our calculations indicate that the chemical potential
vanishes asymptotically as the density tends to zero, without changing
its sign.  The absence of clear evidence of a BEC of dineutrons is the
consequence of the fact that their mutual interaction in free space
does not support a bound state. In other words, the free-space
Schr\"odinger equation for neutrons does not have eigenvalues that
correspond to a dineutron bound state. Even so, it should be
acknowledged that we do find that the ratio of inter-neutron distance
$d$ and the condensate coherence length $\xi_{a}$ satisfies the
conditions $d/\xi_{a}\ll 1$ at high density (in the range under
consideration) and $d/\xi_{a}\ge 1$ at low densities, consistent with
the initial studies~\cite{2006PhRvC..73d4309M,2007PhRvC..76f4316M}.
The values of interparticle spacing $d$ and coherence length $\xi_{a}$
are shown for our model in Table~\ref{table_1} for the case of low
temperature ($T=0.25$ MeV) and vanishing spin polarization at three
values of the density covering the low, intermediate, and high density
regimes.  It is seen that $d/\xi_a \sim 1$ at low densities, which is
a clear sign of a BEC precursor. We address the effects of
polarization on the BCS-quasi-BEC crossover in the following sections.

\subsection{Phase diagram}
\label{subsec:phase_diag}
\begin{figure}[tb]
\label{subsec:gap}
\begin{center}
\vskip 0.7cm
\includegraphics[width=0.46\textwidth,keepaspectratio]{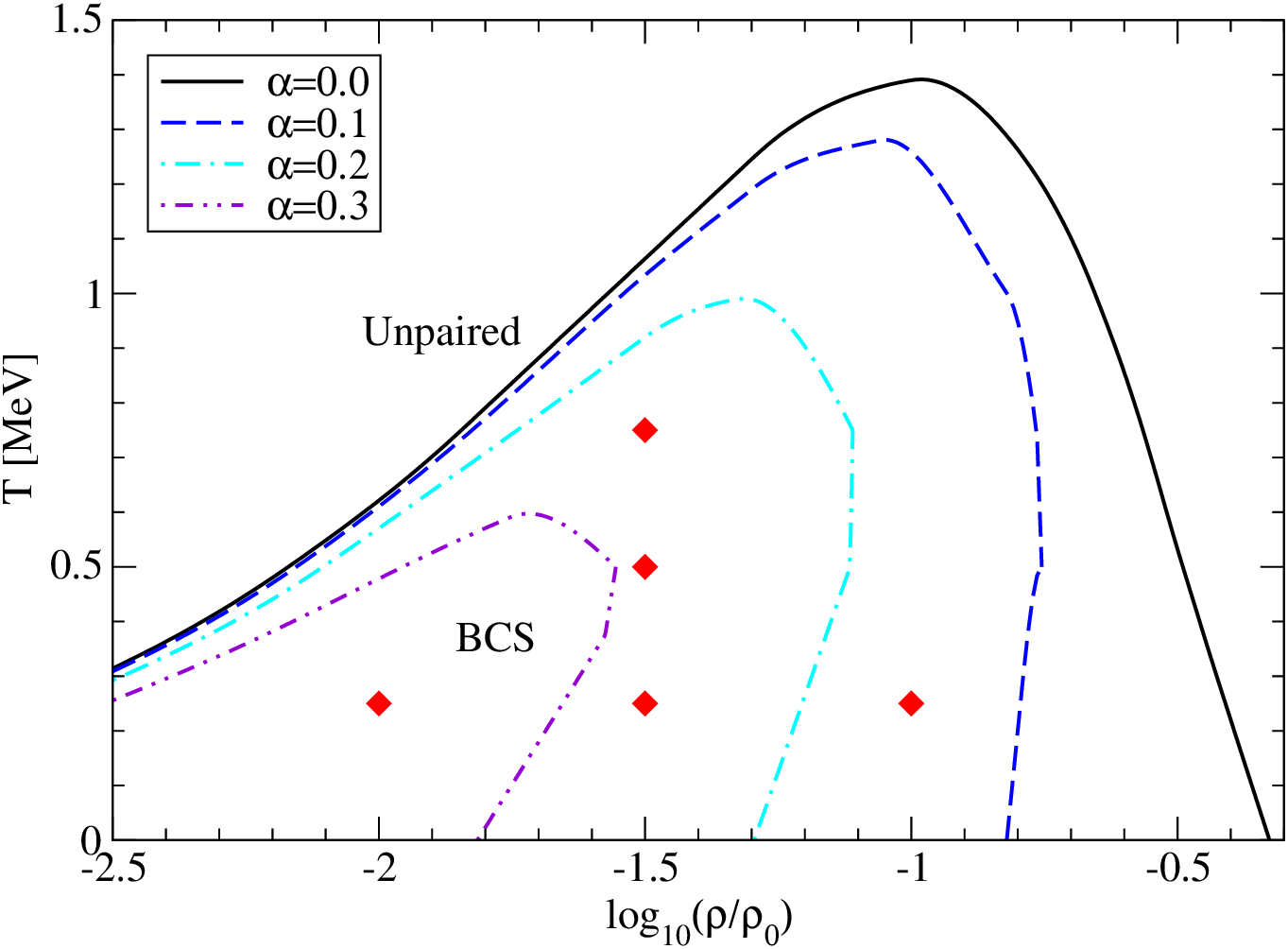}
\caption{(Color online) 
Phase diagram of neutron matter in the temperature-density plane 
for several spin polarizations $\alpha$ induced by magnetic fields. 
The BCS phase is naturally favored over the unpaired phase at lower 
temperatures and smaller polarizations.  The red diamonds locate 
different points in the phase diagram at which some intrinsic features 
of low-density neutron matter have been evaluated.}
\label{phase_diagram_nn}
\end{center}
\end{figure}
The phase diagram was computed by solving Eqs.~\eqref{eq:gap} and
\eqref{eq:densities} self-consistently for the input pairing interaction
in the $^1S_0$ channel. After the solutions were found we evaluated
the free-energy \eqref{eq:free} and found its minimum.  The resulting
phase diagram of neutron matter is shown in
Fig.~\ref{phase_diagram_nn}.  Broadly speaking, we obtain the same
structure as in the case of nuclear matter (cf.\ Fig.~1 of
Ref.~\cite{2012PhRvC..86f2801S}).  At low densities the critical
temperature increases with increasing density, because the increase in
the density of states of neutrons compensates for the decrease in the
attractive interaction strength in the $S$-wave channel with the
increasing Fermi energy of the neutrons. This trend reverses at higher
densities, and the pairing ceases at the point where the interaction
in the $^1S_0$ channel becomes repulsive.  {Spin polarization}
suppresses pairing more efficiently in the high-density sector, where
large portions of the phase diagram are converted from the superfluid
to the normal phase already at moderate spin polarizations.

An interesting feature revealed in Fig.~\ref{phase_diagram_nn} is that
the transition line separating unpaired and BCS phases is not a
single-valued function of density in the range of densities
considered.  This behavior is well understood.  Consider for example
the dot-dash (light blue) transition line in the phase diagram
corresponding the fixed polarization $\alpha = 0.2$.  At low
temperatures and not too low density, pairing is precluded because the
reduced thermal smearing of the Fermi surfaces of the major and minor
spin components cannot provide sufficient phase-space overlap of the
corresponding Fermi quasiparticle distributions. The system remains in
the normal, unpaired phase.  Increasing the temperature at fixed
density and polarization asymmetry increases the smearing effect,
thereby enhancing the overlap enough to restore the BCS phase.  {We
  anticipate that some form of the LOFF phase may fill in the low
  temperature ``pocket'' formed by the inward turn of the phase
  separation boundary (cf. Fig.~1 of
  Ref.~\cite{2012PhRvC..86f2801S}). Note, however, that the pairing
  interaction in neutron matter is weaker than in nuclear matter;
  therefore the stability of the LOFF phase is not guaranteed.

\begin{figure}[b,t]
\label{subsec:gap}
\begin{center}
\vskip 1.cm
\includegraphics[width=0.46\textwidth,keepaspectratio]{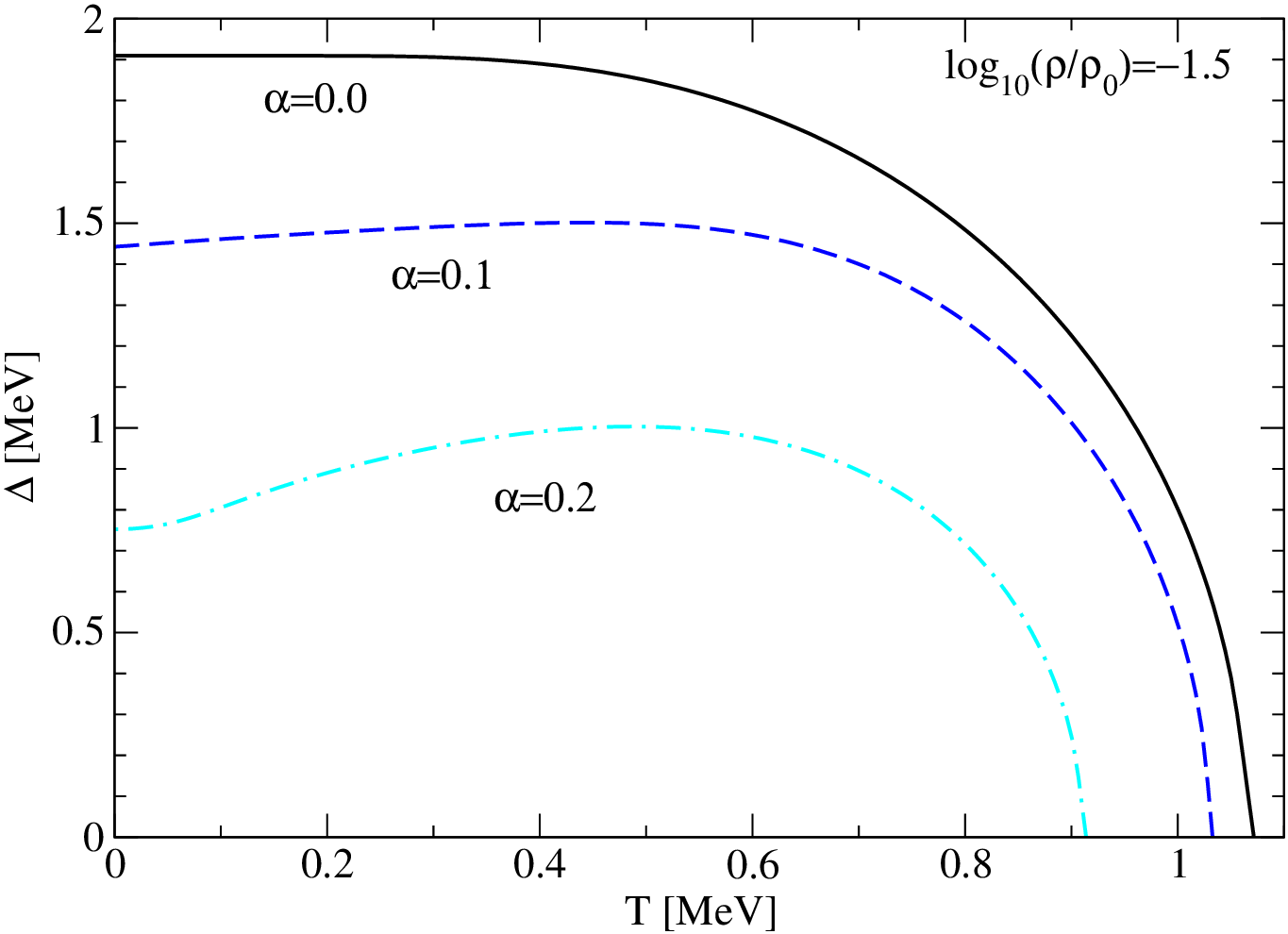}
\caption{(Color online) 
Pairing gap as a function of temperature at constant density 
index $\log_{10}(\rho/\rho_0)=-1.5$ for three choices of polarization 
parameter $\alpha$.}
\label{fig_nn:gap_t}
\hspace{0.1\textwidth}
\vskip 1.cm
\includegraphics[width=0.46\textwidth,keepaspectratio]{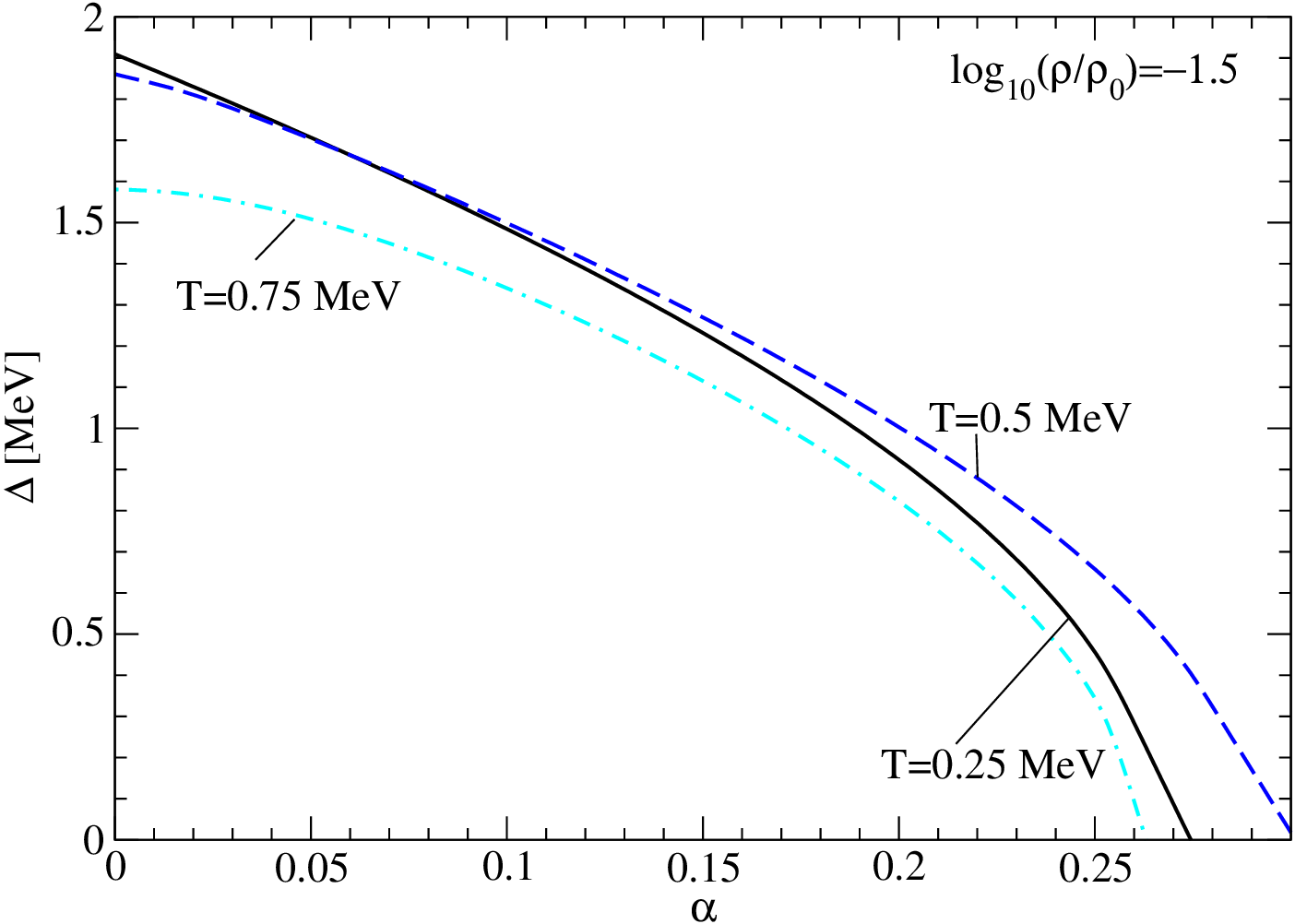}
\caption{(Color online) 
Pairing gap as a function of polarization at constant 
density index $\log_{10}(\rho/\rho_0)=-1.5$ for selected reference 
temperatures.} 
\label{fig_nn:gap_a}
\end{center}
\end{figure}
\subsection{Intrinsic properties of the neutron condensate}

We now proceed to examine some intrinsic features of the isospin-triplet
$^1S_0$ neutron condensate. 
\subsubsection{Pairing gap}

In Figs.~\ref{fig_nn:gap_t} and~\ref{fig_nn:gap_a} we display the gap
at fixed density $\log_{10}(\rho/\rho_0)=-1.5$. In Fig.~\ref{fig_nn:gap_t}
the gap is plotted as a function of temperature for several polarization
values. For zero polarization, i.e., the case of the symmetrical 
BCS state, the value of the gap is maximal owing to perfect
overlap of the Fermi surfaces of up-spin and down-spin particles. The
temperature dependence of the gap corresponds to the standard BCS
behaviour.  Increasing the spin asymmetry has two effects. First, 
the gap is decreased owing to the separation of the Fermi surfaces, and 
so is the critical temperature $T_c$. Second, the maximum of the 
gap is shifted from $T=0$ to nonvanishing temperatures.  For large enough
polarizations, this shift can lead to the appearance of a lower critical 
temperature. 
\begin{figure}[b,t]
\begin{center}
\vskip 1.cm
\includegraphics[width=0.46\textwidth,keepaspectratio]{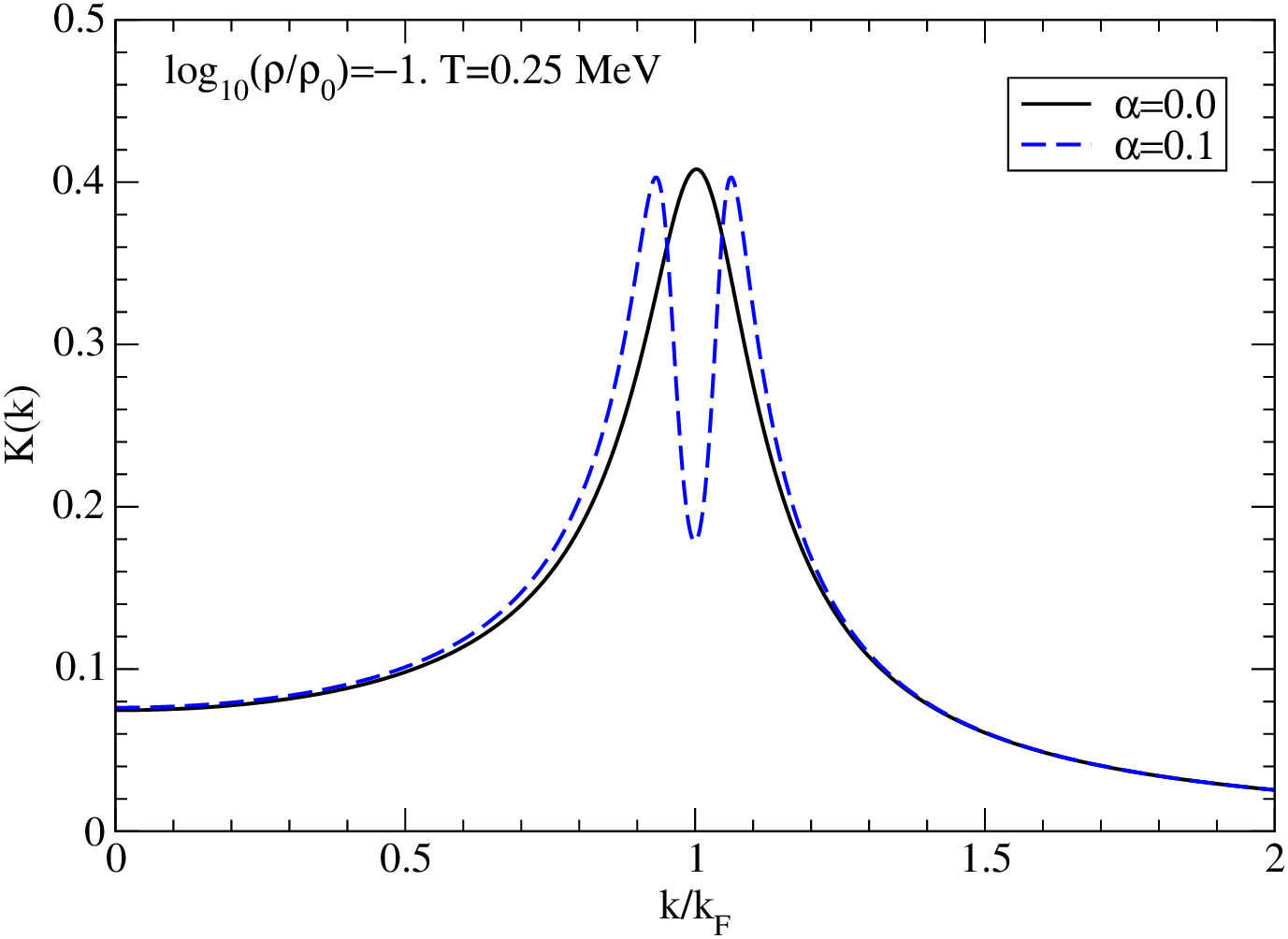}
\caption{(Color online) 
Dependence of the kernel $K(k)$ on momentum (in units of Fermi 
momentum) for fixed $\log_{10}(\rho/\rho_0)=-1$, $T=0.25$ MeV, and polarization 
values color-coded in the inset.}
\label{fig_nn:kernel_1}
\vskip 1.cm
\includegraphics[width=0.46\textwidth,keepaspectratio]{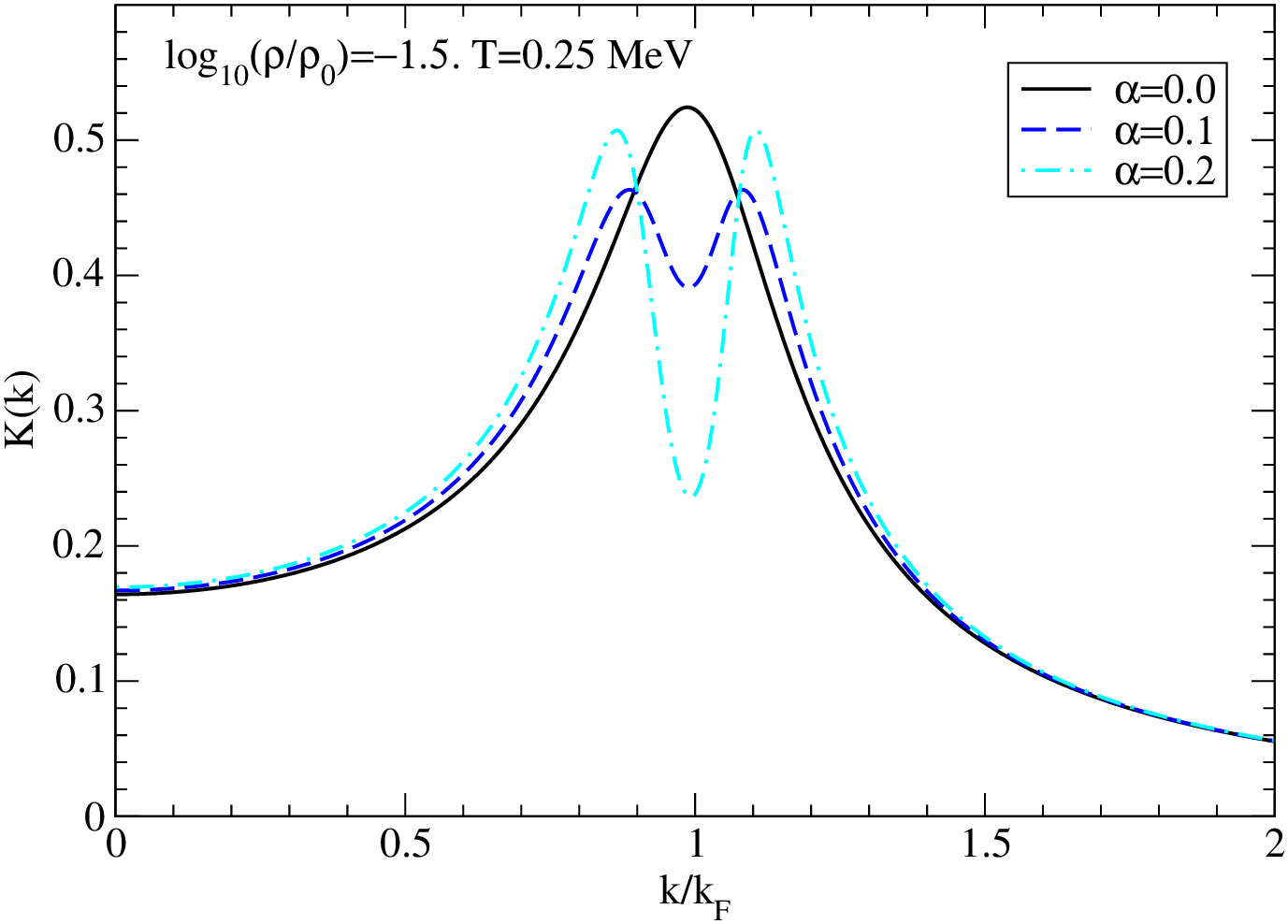}
\caption{(Color online)
Same as Fig.~\ref{fig_nn:kernel_1}, but for $\log_{10}(\rho/\rho_0)=-1.5$ 
and three polarization values color coded in the inset. }
\label{fig_nn:kernel_2}
\end{center}
\end{figure}
\begin{figure}[b,t]
\begin{center}
\vskip 1.cm
\includegraphics[width=0.46\textwidth,keepaspectratio]{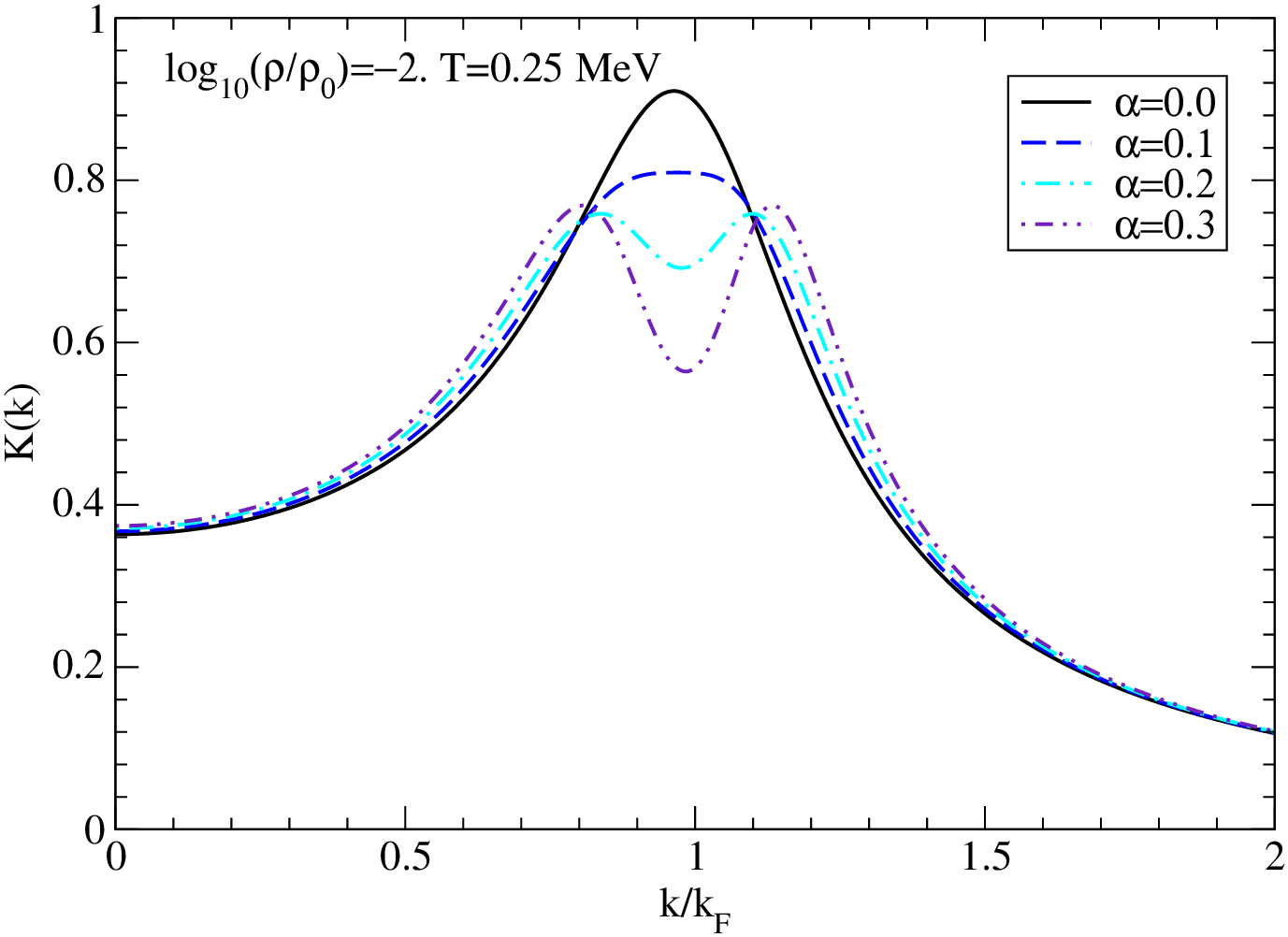}
\caption{(Color online)
Same as Fig.~\ref{fig_nn:kernel_1} but for 
$\log_{10}(\rho/\rho_0)=-2$ 
and more polarization values.}
\label{fig_nn:kernel_3}
\vskip 1.cm
\includegraphics[width=0.46\textwidth,keepaspectratio]{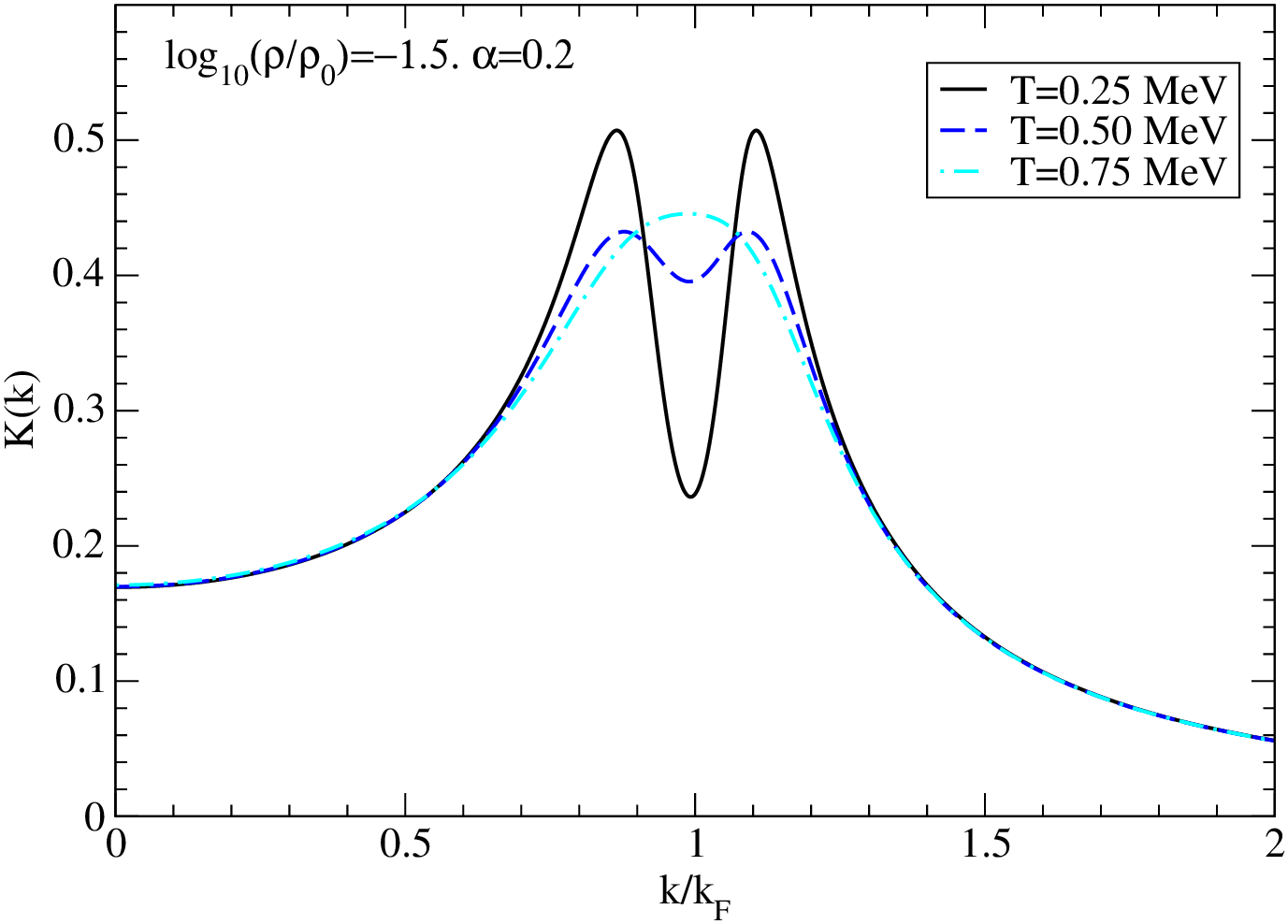}
\caption{(Color online)
Dependence of the kernel $K(k)$ on momentum (in units of 
Fermi momentum) for fixed 
$\log_{10}(\rho/\rho_0)=-1$, $\alpha=0.2$, and 
temperature values color coded in the inset.
}
\label{fig_nn:kernel_4}
\end{center}
\end{figure}
\begin{figure}[b,t]
\begin{center}
\vskip 1.cm
\includegraphics[width=0.46\textwidth,keepaspectratio]{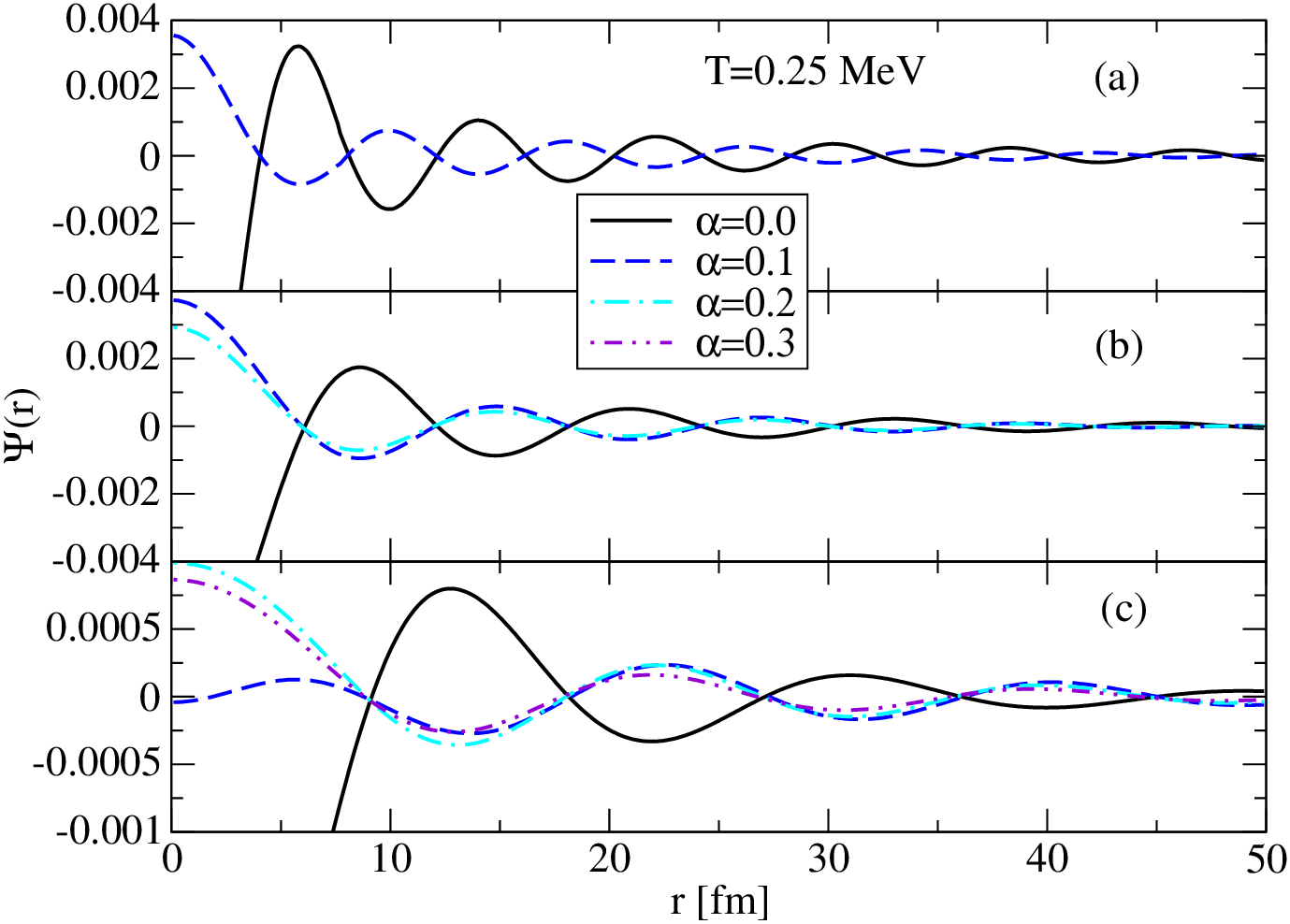}
\caption{(Color online)
Plots of $\Psi(r)$ versus $r$ at fixed temperature $T=0.25$ MeV
for three reference densities $\textrm{log}_{10}(\rho/\rho_0) = -1$
(a), $\textrm{log}_{10}(\rho/\rho_0) = -1.5$ (b), 
and $\textrm{log}_{10}(\rho/\rho_0) = -2$ (c) and 
polarization values $\alpha = 0$ (solid line), $0.1$ (dashed
line),  $0.2$ (dash-dotted), and 0.3 (dashed-double-dotted). }
\label{fig_nn:psi}
\vskip 1.cm
\includegraphics[width=0.46\textwidth,keepaspectratio]{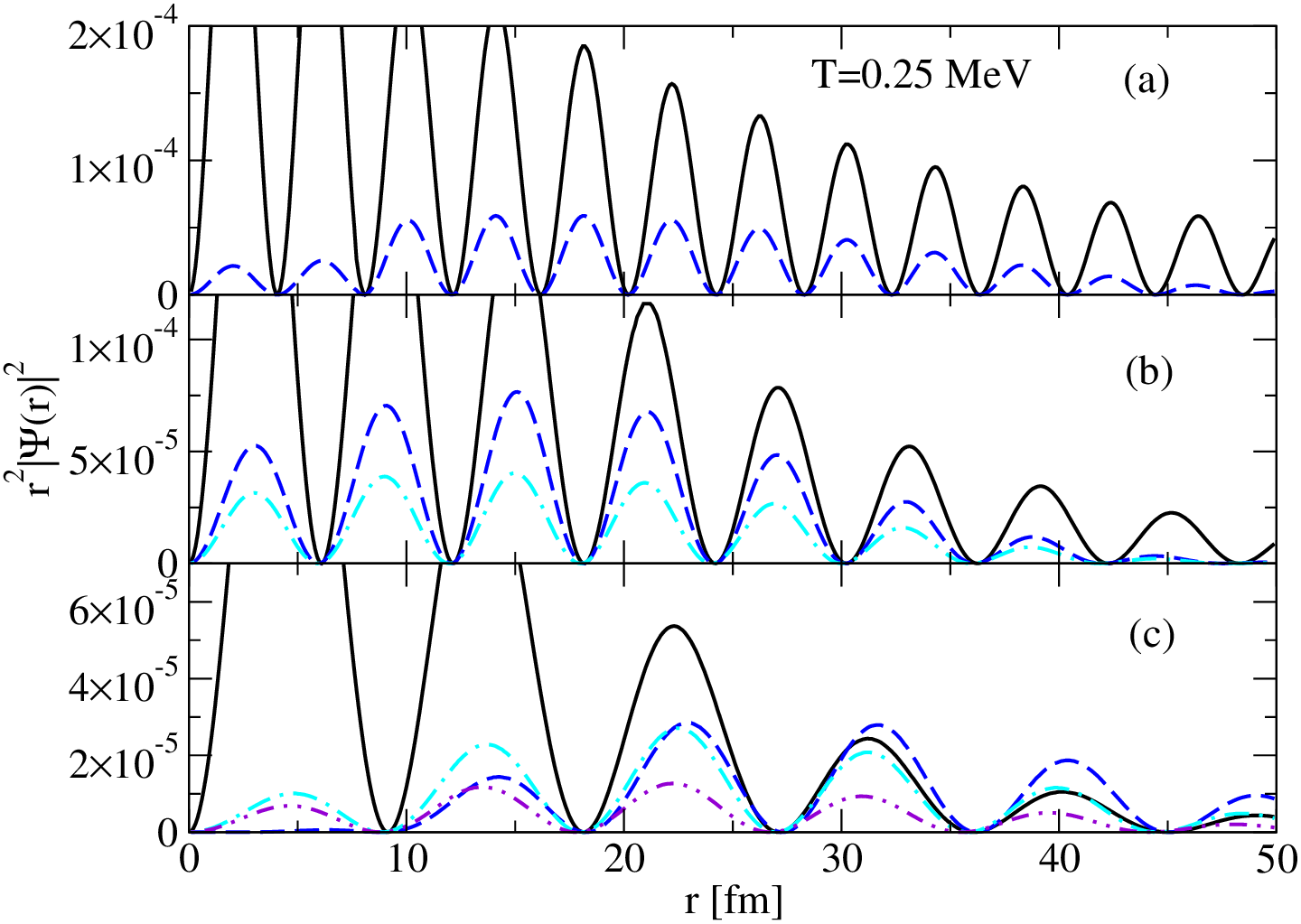}
\caption{(Color online)
Same as Fig.~\ref{fig_nn:psi}, except $r^2|\Psi(r)|^2$ 
is plotted versus $r$.}
\label{fig_nn:rpsi}
\end{center}
\end{figure}
Figure~\ref{fig_nn:gap_a} shows the gap as a function of the
polarization asymmetry parameter $\alpha$ over a range of
temperatures. For $\alpha = 0$, increasing the temperature decreases
the gap, as it should, according to BCS theory. The crossing of
constant-temperature curves at finite $\alpha$ reflects the fact that
raising the temperature from a relatively low value favors pairing in
asymmetrical systems, by virtue of the increased overlap between the
Fermi surfaces of the opposite-spin components. Of course, at
high-enough temperatures this effect must give way instead to the
destruction of the superconducting state.  These competing effects are
reflected in the Fig.~\ref{fig_nn:gap_a}.  At high-enough
polarizations, the increase of temperature from $T=0.25$ MeV to
$T=0.5$ MeV increases the gap, whereas the increase of temperature
from $T=0.5$ MeV to $T=0.75$ MeV acts to reduce the gap.  {Note that
  allowing for the LOFF phase will modify the low-temperature behavior
  seen in Figs.~\ref{fig_nn:gap_t} and~\ref{fig_nn:gap_a} in a
  well-known
  manner~\cite{2012PhRvC..86f2801S,2014PhRvC..90f5804S,2006PhRvB..74u4516H}.

\subsubsection{Kernel of the gap equation}
In Figs.~\ref{fig_nn:kernel_1}--\ref{fig_nn:kernel_4} we present the
kernel of the gap equation for various values of density, temperature,
and polarization in the BCS phase.  The kernel of the gap
equation is defined as 
\begin{eqnarray}
K(k) &=& \frac{1}{4}\sum_{a,r}\frac{1-2f(E^a_r)}{\sqrt{E_{S}^2(k)+\Delta^2(k)}}.
\end{eqnarray}
Figures~\ref{fig_nn:kernel_1}--\ref{fig_nn:kernel_3} show the kernel at
$T=0.25$ MeV for several values of the polarization, the density being
fixed for each figure. As expected in the case of $\alpha=0$ we find a
single peak centered at the Fermi level. This peak separates into two for
nonvanishing polarizations, simply reflecting the fact that there are now
the two Fermi surfaces for up-spin and down-spin particles.  In these
figures one also observes that at high densities the peak of the kernel is 
located exactly at $k=k_F$, whereas for low densities the peak 
is shifted to momenta below the corresponding $k_F$. Additionally,
at lower densities the polarization-induced two-peak structure is 
smeared; this is naturally attributed to the weakening of the degeneracy 
of the system.

The kernel evaluated at constant density and polarization is exhibited in 
Fig.~\ref{fig_nn:kernel_4} for three different temperatures.  One clearly
recognizes a thermal smearing of the polarization-induced two-peak
structure, which evolves into a one-peak structure at high temperatures.

\subsubsection{Cooper-pair wave function}

Next we discuss the Cooper-pair wave function $\Psi(r)$ and the 
quantity $r^2\vert\Psi(r)\vert^2$, which determines the second moment of
the density distribution of Cooper pairs. With the wave function at 
our disposal, we also have numerical access to the correlation  
length $\xi_\mathrm{rms}$ of the condensate, which can then
be compared with the analytical BCS expression for the coherence 
length $\xi_a$ and with the interparticle distance $d$.  The wave 
function is obtained by the Fourier transformation as 
\bea \Psi(r)&=& \frac{\sqrt{N}}{2\pi^2r} \int_0^\infty
dp\,p\,[K(p,\Delta)-K(p,0)]\sin(pr)\,, \nonumber\\
\eea 
with normalization satisfying
\bea 1&=&N\int d^3r \vert \Psi(r)\vert^2.  
\eea 
The root-mean-square (rms) value for the coherence length is given by 
\bea
\xi_{\rm rms} &=& \sqrt{\langle r^2\rangle}\,, 
\eea 
where 
\bea \langle
r^2\rangle &\equiv& \int d^3r\, r^2 \vert\Psi(r)\vert^2.  
\eea 
The analytical BCS result for the coherence length is given by 
\bea \xi_a &=&
\frac{\hbar^2 k_F}{\pi m^* \Delta}\,, 
\eea 
where now $\Delta$ is the
pairing gap in the $^1S_0$ channel and $m^*$ is the effective mass of
neutrons.  Finally, the interparticle distance is simply related to the
total particle density of the system by
\bea d &=&
\left(\frac3{4\pi\rho}\right)^{1/3}.  
\eea 

Table~\ref{table_1} displays the quantities defined above at vanishing
polarization and fixed temperature $T=0.25$ MeV.  For each of three 
representative densities, corresponding values are entered for
$k_F$, $\Delta$, $m^*/m$, $\mu_n$, $d$, $\xi_\mathrm{rms}$, and 
$\xi_a$.  At high density it is seen that $\xi_\mathrm{rms}\simeq \xi_a$, 
i.e., the BCS analytical expression is a good approximation to the 
numerically computed coherence length. This is not the case at low
densities, where one can only rely on the numerical value $\xi_\mathrm{rms}$ 
produced by our theoretical treatment.  At any rate, comparison of
the numerically generated coherence length with the interparticle
distance shows a clear signature of a BCS-BEC crossover: For
$\log_{10} (\rho/\rho_0) = -1$ we find $\xi_\mathrm{rms}/d \simeq 2$,
whereas for $\log_{10} (\rho/\rho_0) = -2$ the pertinent ratio is
$\xi_\mathrm{rms}/d \simeq 0.45$.  Below we will trace, in other 
variables, further signatures of a BCS-BEC crossover in spin-polarized
neutron matter.

In Fig.~\ref{fig_nn:psi} the Cooper-pair wave function $\Psi(r)$ is plotted
against radial distance at fixed temperature $T=0.25$ MeV and various 
polarization values, for the three fiducial densities adopted in Table I.  
In all cases we find strongly oscillating wave functions. For nonvanishing 
polarization, the wave function experiences a sign change; the
oscillations are then in counterphase to the unpolarized case. 
With increasing polarization, the amplitude of $\Psi(r)$ decreases 
in accord with the consequent reduction of the pairing gap. Furthermore,
as the oscillation periods are given roughly by $2\pi/k_F$, a decrease 
of density and hence of Fermi momentum leads to an increase of oscillation
period.  The degree of polarization does not affect the period, which 
is determined by $k_F$ values.  Figure~\ref{fig_nn:rpsi} shows
$r^2\vert\Psi(r)\vert^2$ as a function of radial distance, 
the oscillatory behavior observed in Fig.~\ref{fig_nn:psi} being
reflected quite naturally in this quantity.  However, two features are
made more apparent in Fig.~\ref{fig_nn:rpsi}. At the lowest density
considered, (i) the maxima of this wave function measure plotted for 
different polarizations are shifted with respect to each other
and (ii) the overall maximum attained for each polarization is not
situated at the same value of $r$ (although this does become the 
case at higher densities).

\begin{figure}[b,t]
\begin{center}
\vskip 1.cm
\includegraphics[width=0.46\textwidth,keepaspectratio]{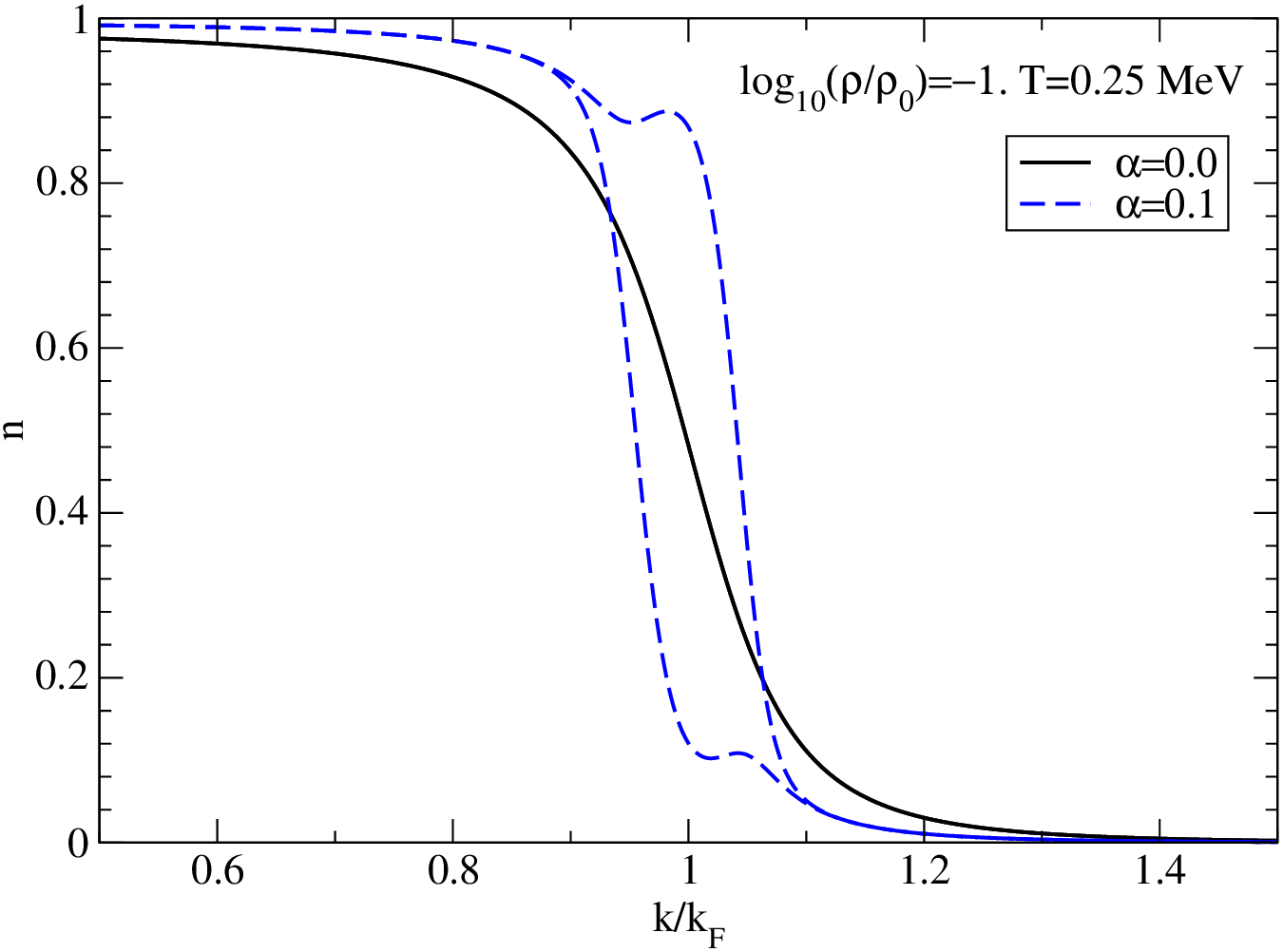}
\caption{(Color online)
Dependence of the up-spin and down-spin neutron occupation numbers 
on momentum $k$ (in units of Fermi momentum) for fixed $\log_{10}(\rho/\rho_0)=-1$, 
$T=0.25$ MeV, and polarization values color coded in the inset.}
\label{fig_nn:occupation_1}
\vskip 1.cm
\includegraphics[width=0.46\textwidth,keepaspectratio]{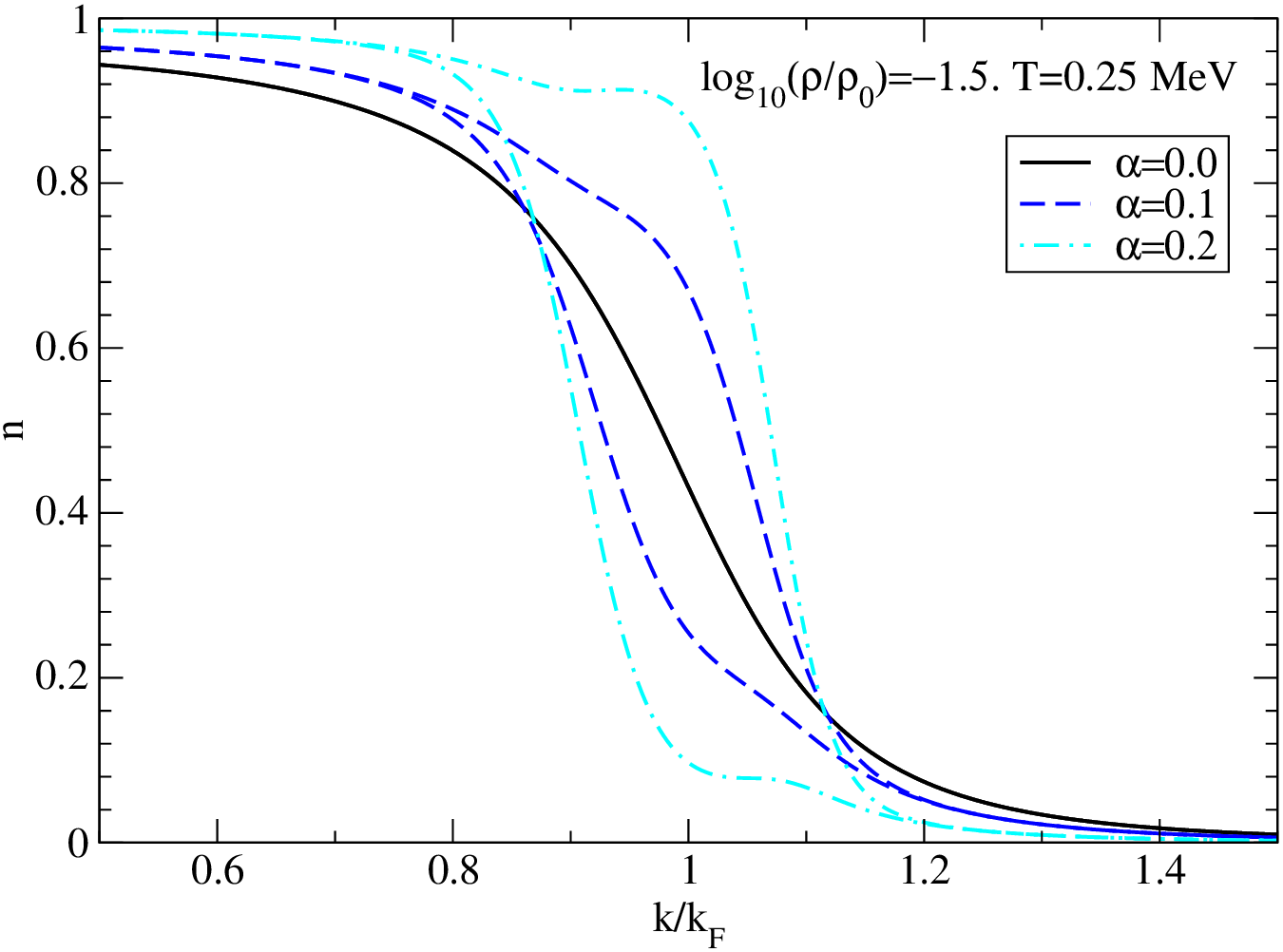}
\caption{(Color online)
Same as Fig.~\ref{fig_nn:occupation_1}, but for $\log_{10}(\rho/\rho_0) 
=-1.5$ and an additional polarization value.}
\label{fig_nn:occupation_2}
\end{center}
\end{figure}

\begin{figure}[b,t]
\begin{center}
\vskip 1.cm
\includegraphics[width=0.46\textwidth,keepaspectratio]{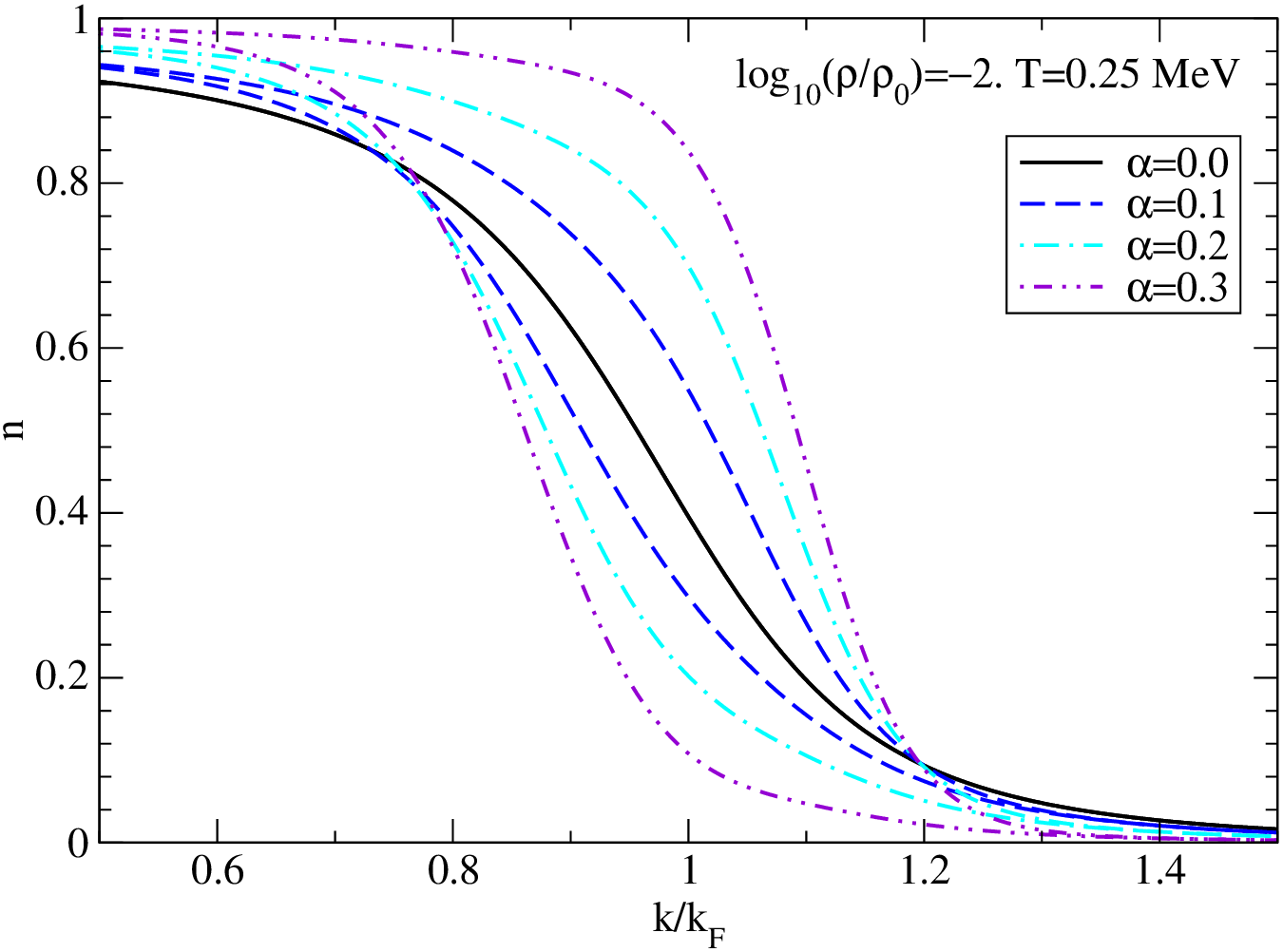}
\caption{(Color online)
Same as Fig.~\ref{fig_nn:occupation_2}, but for $\log_{10}(\rho/\rho_0)=-2$ 
and an additional polarization value.}
\label{fig_nn:occupation_3}
\vskip 1.cm
\includegraphics[width=0.46\textwidth,keepaspectratio]{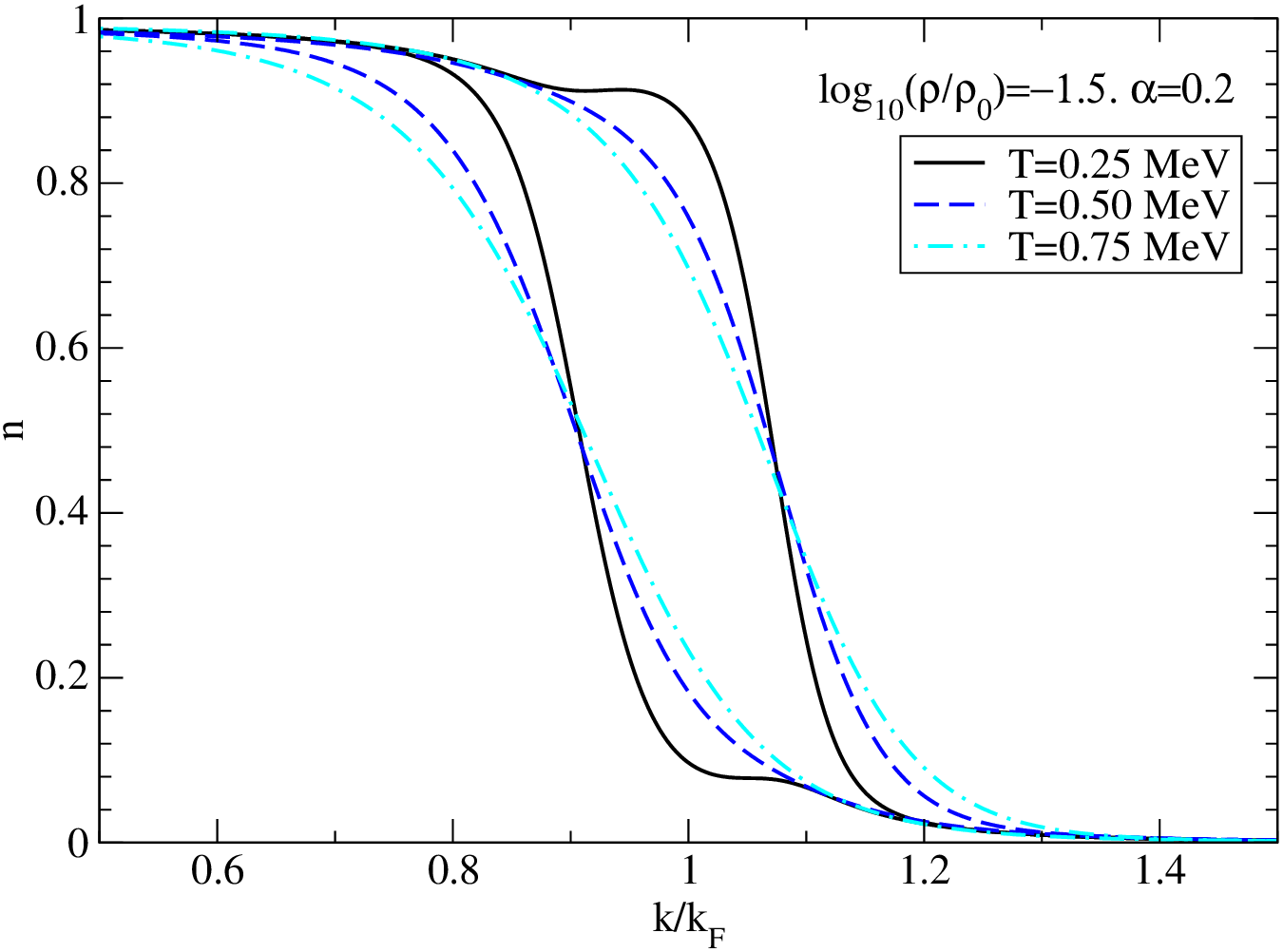}
\caption{(Color online)
Dependence of the up-spin and down-spin neutron occupation numbers 
on momentum $k$ (in units of Fermi momentum) for fixed $\log_{10}(\rho/\rho_0) 
=-1.5$, $\alpha=0.2$, and temperatures color coded in the inset.} 
\label{fig_nn:occupation_4}
\end{center}
\end{figure}
\begin{figure}[b,t]
\begin{center}
\includegraphics[width=0.46\textwidth,keepaspectratio]{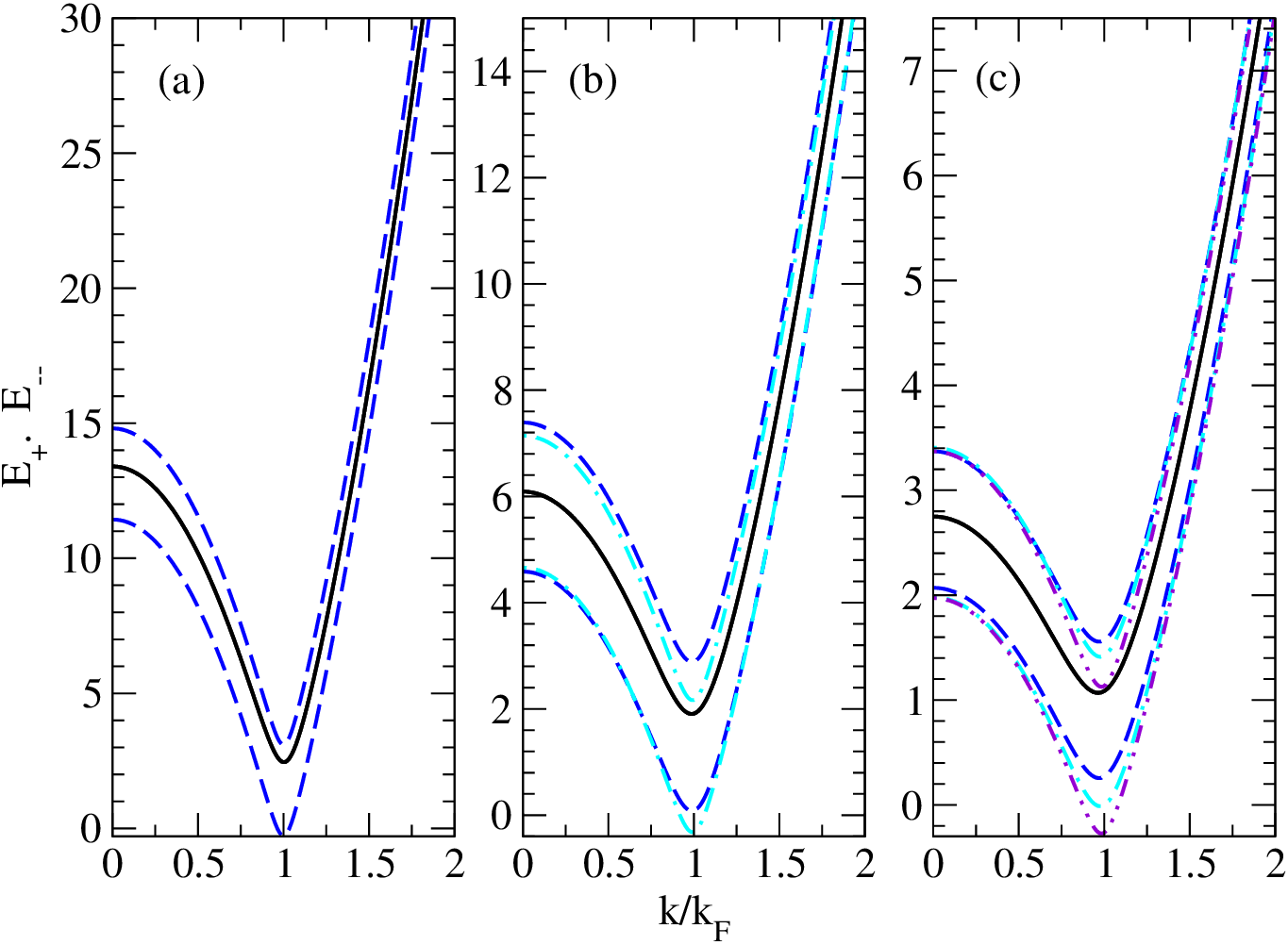}
\caption{(Color online) Dispersion relations for quasiparticle spectra 
  in the BCS condensate, as functions of momentum (in units of Fermi 
  momentum) for three densities $\textrm{log}_{10}(\rho/\rho_0) = -1$
  (a), $\textrm{log}_{10}(\rho/\rho_0) = -1.5$ (b), and 
  $\textrm{log}_{10}(\rho/\rho_0) = -2$ (c).  The polarization 
  values are $\alpha = 0$ (solid line), $0.1$ (dashed line), $0.2$
  (dash-dotted), 0.3 (dashed-double-dotted).  For each polarization,
  the upper branch corresponds to the $E_+$ solution, the lower branch 
  to the $E_-$ solution.  }
\label{fig_nn:quasiparticle}
\end{center}
\end{figure}
\subsubsection{Quasiparticle occupation numbers}

In this section, we analyze the behavior of the occupation numbers 
of up-spin and down-spin neutrons in spin-polarized pure neutron matter. The
occupation numbers are given by the integrand of Eq.~\eqref{eq:densities}. 
Explicitly,
\begin{eqnarray}
  n_{\uparrow/\downarrow}(k)
  &=&\frac12\left(1+\frac{E_S}{\sqrt{E_S^2+\Delta^2}}\right)f(E_\mp)\nonumber\\
  &+&\frac12\left(1-\frac{E_S}{\sqrt{E_S^2+\Delta^2}}\right)
[1-f(E_\pm)]\,,\label{eq_n_nn}
\end{eqnarray}
with $E_r^a\rightarrow E_r$ for BCS pairing with $\vecQ=0$, i.e.,
$E_A = 0$ in Eq.~(\ref{eq:E_A}). It may be noted in passing that 
the functions $ n_{\uparrow/\downarrow}(k)$ have maximum value 1, 
rather than the value 2 appropriate to nuclear matter (which 
reflects a summation over spin).

Figures~\ref{fig_nn:occupation_1}--\ref{fig_nn:occupation_3} 
display the occupation numbers of up-spin and down-spin neutrons at fixed
temperature $T=0.25$ MeV and fixed densities $\log_{10} (\rho/\rho_0) = -1$,
$-1.5$, and $-2$ respectively. The chosen polarization values are indicated
in each figure.  We observe that in the case of vanishing polarization 
(solid lines), the Fermi-step-like occupation present in the high-density 
limit has evolved into an increasingly flatter distribution at low 
densities, the Fermi surface growing ever more diffuse with decreasing
density.  At finite polarizations, the occupation numbers (or occupation
probabilities) of up-spin and down-spin neutrons ``split,'' or separate 
from one another, into distinct curves in the region around $k_F$.
In fact, the locations of the drop-offs in the occupancies of these 
two spin populations agree well with their corresponding Fermi wave 
numbers. At high densities, the polarization-induced splitting
results in a ``breach'' for large asymmetries with $n_{n\uparrow}\approx 1$ and
$n_{n\downarrow}\approx 0$ around $k_F$. (The notion of  breach
and ``breached pairing'' in the same context was introduced for ultracold atoms in
Ref.~\cite{2003PhRvL..91c2001G}).  The breach remains intact at lower
densities, but the slope of the corresponding occupation probabilities
declines, as already remarked for the case of unpolarized matter. 

In Fig.~\ref{fig_nn:occupation_4} we show the occupation numbers of up-spin
and down-spin neutrons at fixed density $\log_{10}(\rho/\rho_0)=-1.5$ and
fixed polarization $\alpha=0.2$ for different temperatures. As clearly
seen, the occupation probabilities are subjected to greater smearing 
with increasing temperature.

\subsubsection{Quasiparticle spectra}

Turning to the final intrinsic property of interest, we examine the 
dispersion relations for quasiparticle excitations about the $^1S_0$ pairing
condensate. Because a LOFF phase does not enter the picture here, 
the quasiparticle branches $E_{\pm}^{-}$ and $E_{\pm}^{+}$ coincide 
and the superscript may be dropped, leaving just two branches
\begin{eqnarray}
E_{\pm}(k) &=& \sqrt{\left(\frac{k^2}{2m^*}-\bar\mu\right)^2+\Delta^2}\pm \delta\mu .
\end{eqnarray}
These dispersion relations are plotted in
Fig.~\ref{fig_nn:quasiparticle} for various values of density and
polarization at fixed temperature $T=0.25$ MeV. In each case 
the spectrum has a minimum at $k_F$. At finite polarization there 
is a splitting of the spectra of up-spin and down-spin neutrons. 
A special feature that deserves notice is that at low densities 
the spectrum of the minority component (e.g., the down-spin neutrons)
crosses zero, which implies that its spectrum is gapless.

\begin{figure}[tb]
\begin{center}
\includegraphics[width=0.42\textwidth,keepaspectratio]{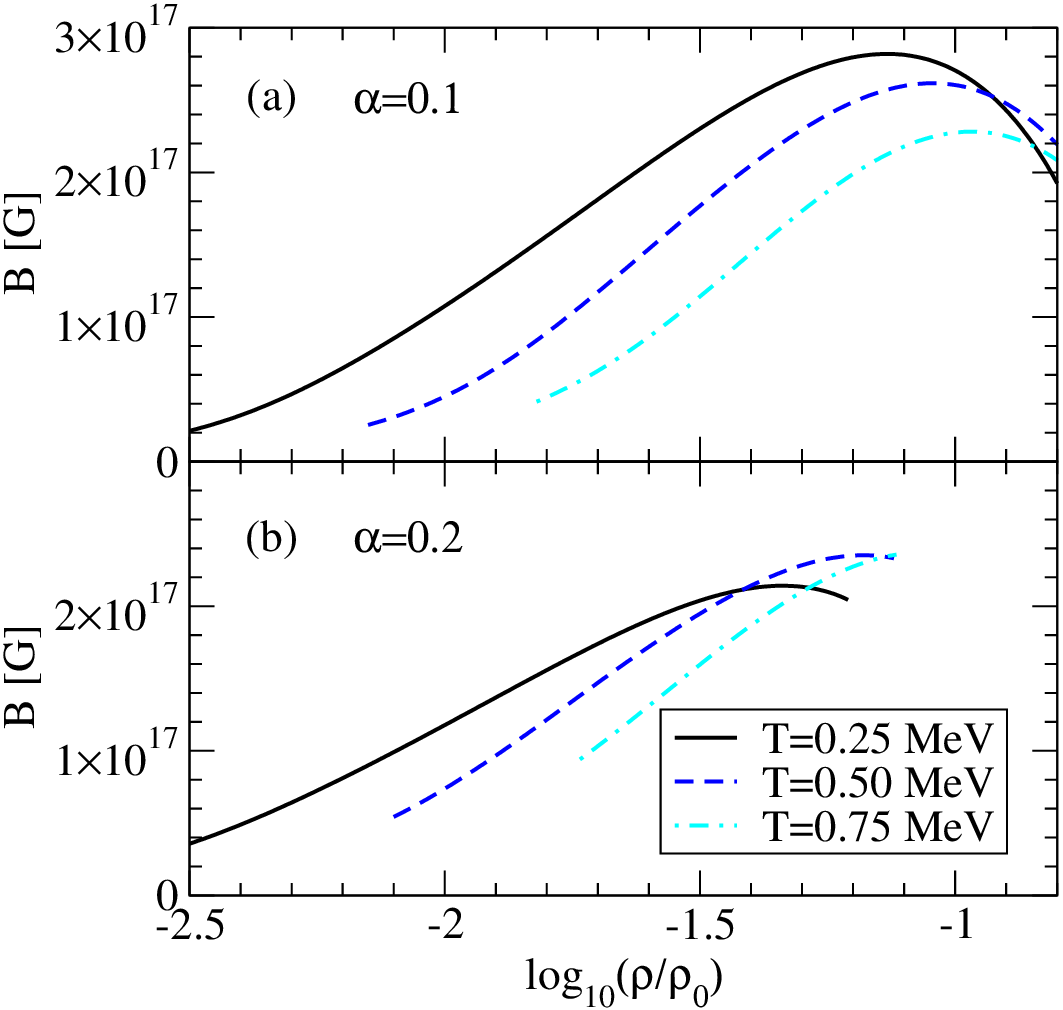}
\caption{(Color online) 
Magnetic field required to create a specified spin polarization 
as a function of the density for two polarization values $\alpha =
0.1$ (a) and 0.2 (b) and temperatures $T=0.25$ (solid 
line),  0.5 (dashed line), and 0.75 (dash-dotted line). 
}
\label{fig_nn:b_field_1}
\includegraphics[width=0.42\textwidth,keepaspectratio]{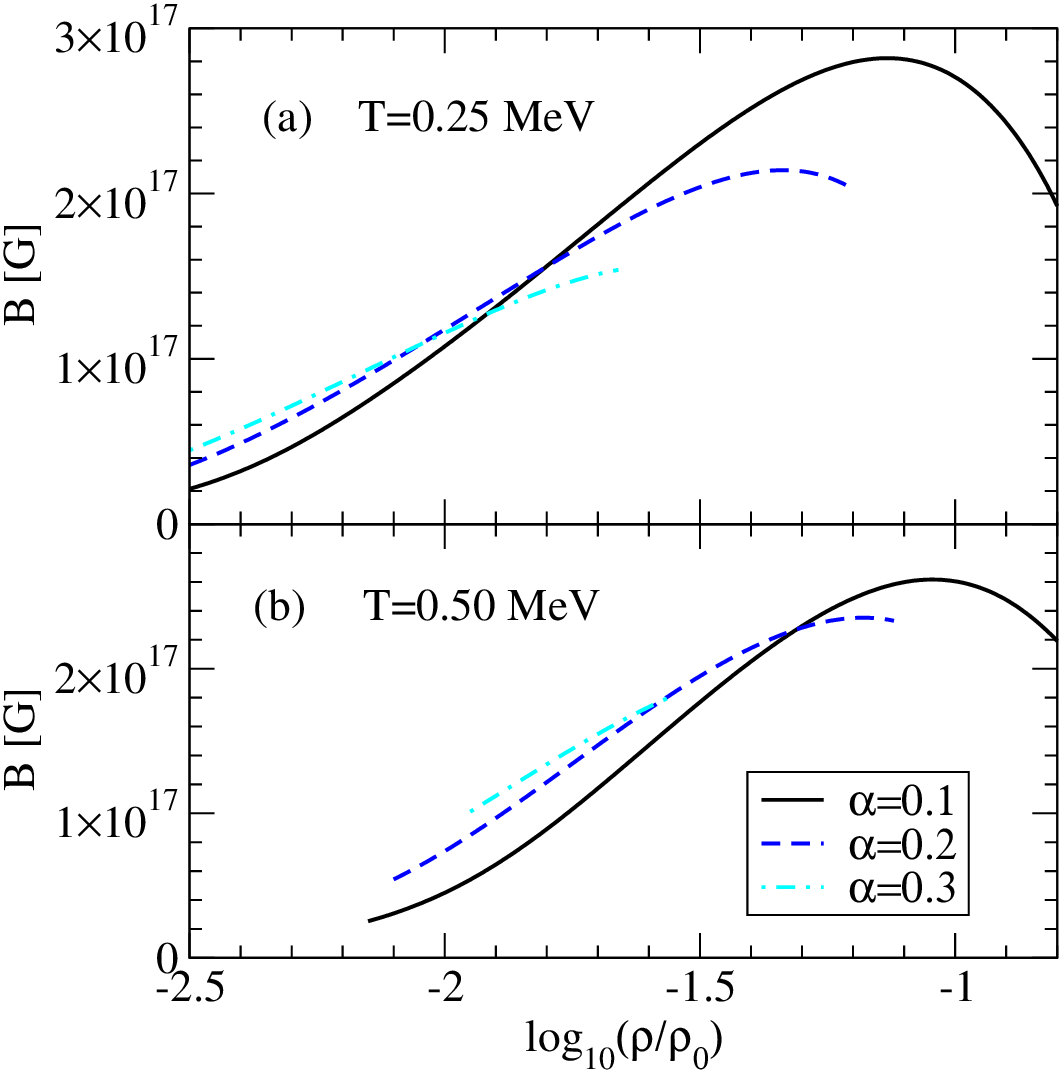}
\caption{(Color online) Same as Fig.~\ref{fig_nn:b_field_1} for two
  temperatures $T=0.25$ MeV (a) and 0.5 MeV (b) and for several
  polarizations $\alpha = 0$ (solid line), 0.2 (dashed line), 0.3
  (dash-dotted line).  }
\label{fig_nn:b_field_2}
\end{center}
\end{figure}
\begin{figure}[tb]
\begin{center}
\includegraphics[width=0.4\textwidth,keepaspectratio]{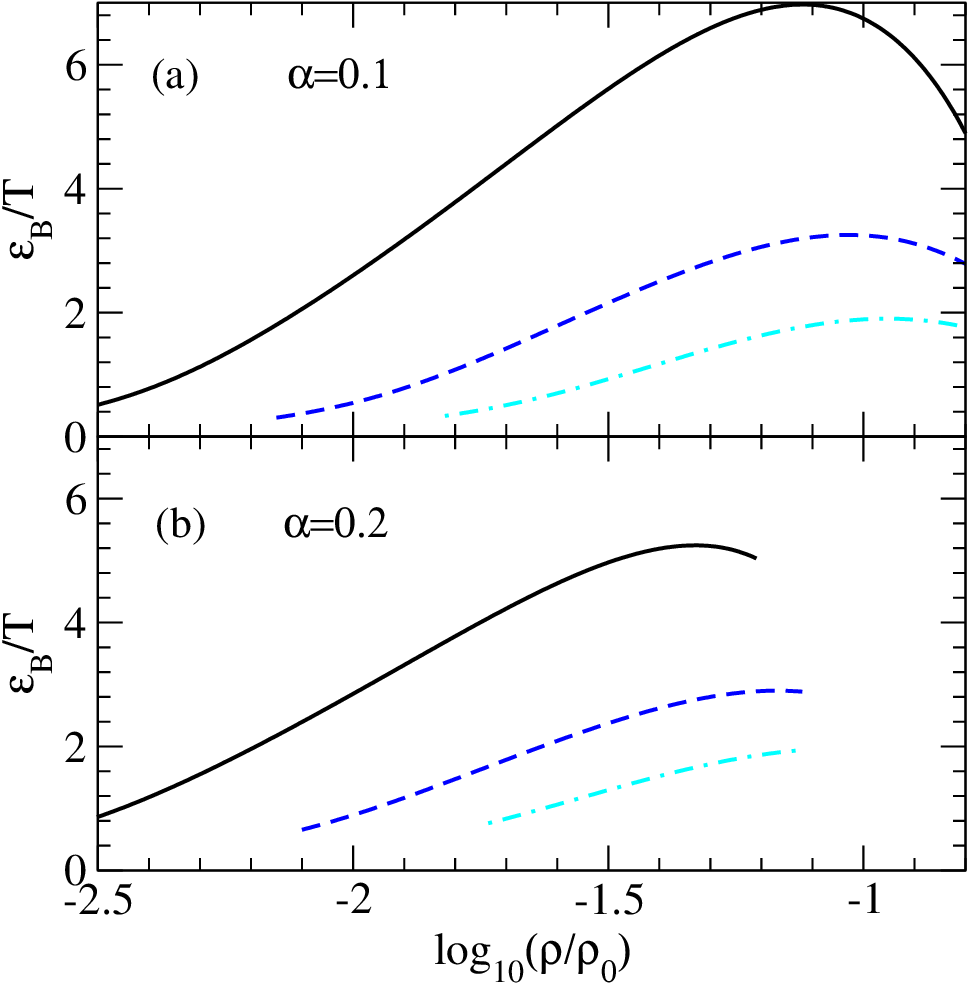}
\caption{(Color online) Ratio of magnetic energy to temperature as a
  function of density for two polarization values $\alpha = 0.1$ (a)
  and 0.2 (b) and temperatures $T=0.25$ (solid line), 0.5 (dashed
  line), and 0.75 (dash-dotted line).  }
\label{fig_nn:b_field_3}
\includegraphics[width=0.4\textwidth,keepaspectratio]{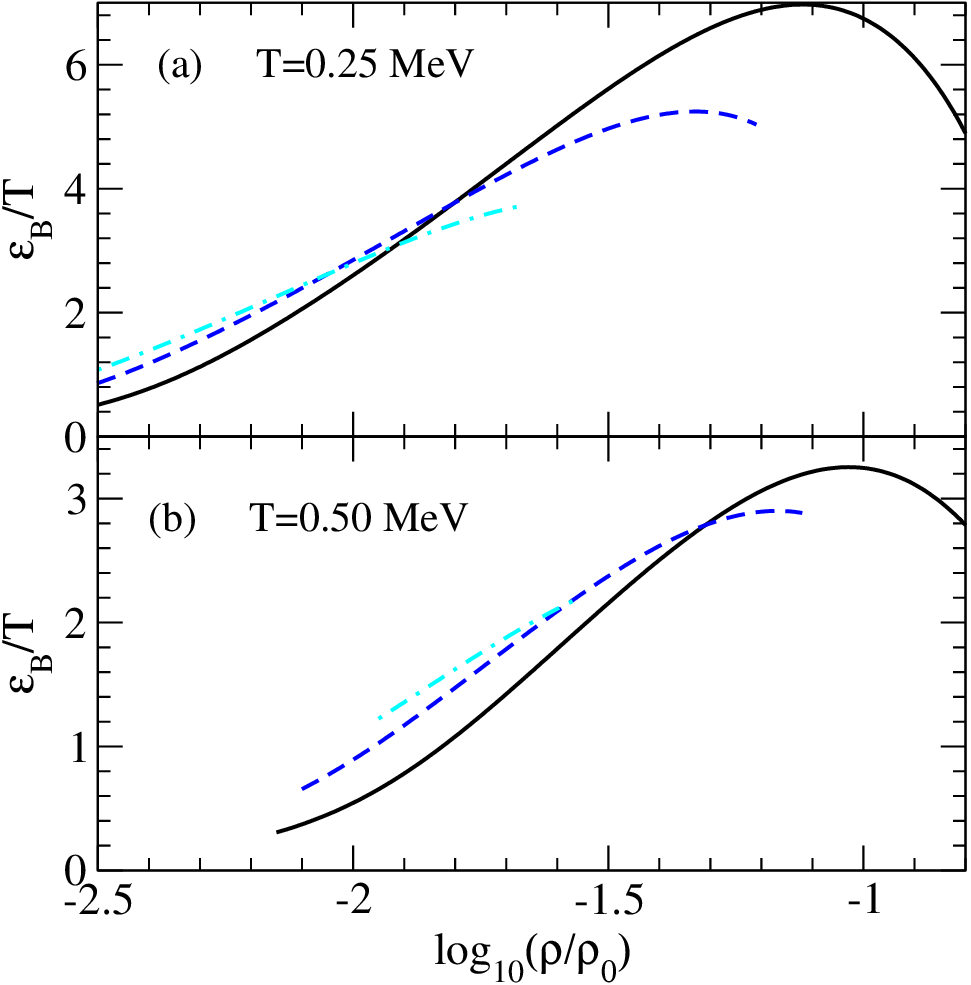}
\caption{(Color online) Ratio of magnetic energy to temperature as a
  function of density for two temperatures $T=0.25$ MeV (a) and 0.5
  MeV (b) and for several polarizations $\alpha = 0$ (solid line), 0.2
  (dashed line), 0.3 (dash-dotted line).  }
\label{fig_nn:b_field_4}
\end{center}
\end{figure}

\section{Critical unpairing in neutron matter}
\label{sec:magentars}
It is elementary that spin polarization in pure neutron matter can be
induced by a magnetic field.  A given polarization corresponds to
shifts having equal magnitude $|\delta \mu|$ xof the chemical potentials 
$\mu_{\uparrow}$ and $\mu_{\downarrow}$ of the up-spin and down-spin
components relative to their common chemical potential at zero
polarization.  The required field magnitude is then given by
\begin{equation}
|\delta \mu| =  \vert \tilde {\mu}_N \vert B ,
\end{equation}
where
\begin{equation}
\tilde \mu_N =  g_n \frac{m_n}{m_n^*} \mu_N
\end{equation}
is the spin magnetic moment of the neutron, with $g_n = -1.91$ its
$g$ factor and $m^*$ its effective mass, $\mu_N= e \hbar /2mc$ being
the nuclear magneton (in cgs units).  Thus, the magnetic field 
involved is linearly related to the shift of chemical potentials 
for a specified spin polarization.

\begin{figure}[b,t]
\includegraphics[width=0.46\textwidth,keepaspectratio]{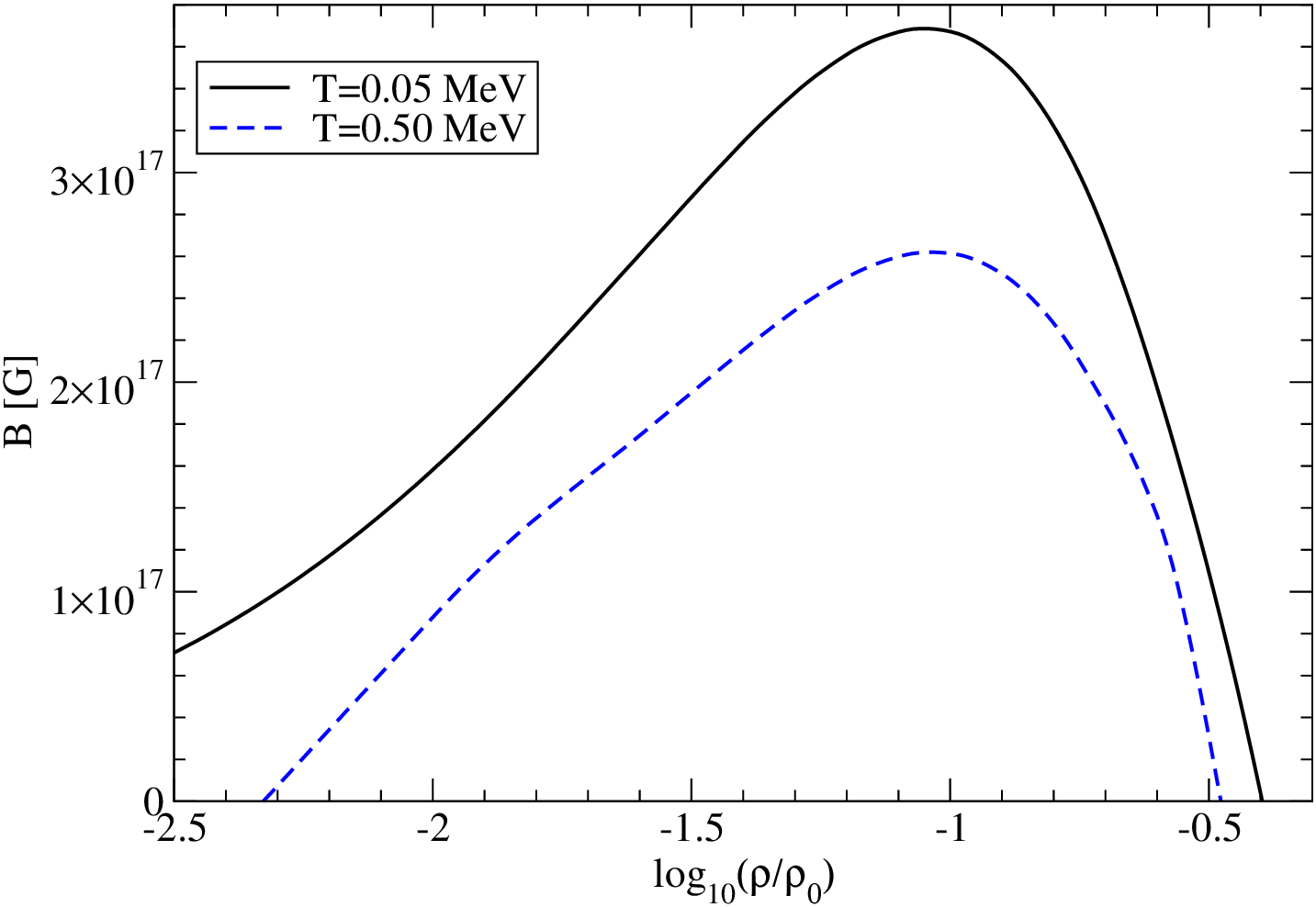}
\caption{(Color online)
Unpairing magnetic field as a function of density (in 
  units of $\rho_0$) for $T = 0.05$ (solid line) 
  and $T = 0.5$ MeV (dashed line). 
}
\label{fig:Bcrit}
\end{figure}

In Fig.~\ref{fig_nn:b_field_1} we display results for the requisite 
magnetic field as a function of density at constant polarization and
temperature.  In the main, this figure tells us that a larger magnetic
field is needed to obtain a given spin polarization as the density
increases.  In other words, dense neutron matter is harder to polarize
than low-density neutron matter.  However, this trend reverses at and
above approximately one-tenth the saturation density $\rho_0$.  The
underlying physical content of this observation is difficult to access
because the chemical potential shift is non trivially related to both
the polarization (the density asymmetry of up-spin and down-spin
components) and the pairing gap.  It is further seen from
Figure~\ref{fig_nn:b_field_1} that higher-temperature neutron matter is
more easily polarizable at low densities, but this trend may again
reverse at higher densities.  Fig.~\ref{fig_nn:b_field_2} provides an
alternative view of the same information, now with the temperature
fixed in each panel and lines of fixed polarization color coded.  From
this view we infer that a larger magnetic field is needed to achieve a
larger polarization in low-density neutron matter.  However, this
trend may again reverse at higher densities.

It is conceptually interesting to examine the ratio of the magnetic
energy  {(associated with the neutron spin's interaction with
  the magnetic field) to the temperature, i.e., the ratio}
 \bea 
\frac{\varepsilon_B}{T} \simeq \frac{\vert \tilde \mu_N\vert  B}{T} .  
\eea 
 {For degenerate neutrons the transport and
  radiation processes involve neutrons located in the narrow strip
  of width $\sim T$ around the Fermi surface; the magnetic field
  influences these processes when this ratio becomes of 
  the order unity. 
}
It is shown in Figs.~\ref{fig_nn:b_field_3} and
\ref{fig_nn:b_field_4}, where the arrangement of the panels and the
color coding are analogous to those of Figs.~\ref{fig_nn:b_field_1} and
\ref{fig_nn:b_field_2}, respectively.  It is seen that
$\epsilon_B/T \gg 1$ over almost the complete range of the parameter space,
with exceptions only at very low densities.  {This
 implies that the dynamical processes in neutron fluid will be
 strongly affected by the field if  the blocking effect of the
 pairing gap can be surmounted, e.~g., when $\Delta(T,\alpha)\le T$.}

Finally, we consider the critical magnetic field that completely
destroys the $^1S_0$ Cooper pairs in neutron matter by aligning
the neutron spins in each pair.  As a function of density, this
field, shown in Fig.~\ref{fig:Bcrit}, has a shape that naturally 
reflects the corresponding density dependence of the pairing
field. Accordingly, it is largest at $T\to 0$ and decreases as 
the pairing gap decreases with increasing temperature.

The strengths of the $B$ fields in the crust and outer-core regions of
magnetars are unknown, although it is anticipated that their interior
fields could be much larger than the surface fields $B\sim 10^{15}$~G
inferred from observations. A number of magnetar models entertain the
possibility that strong toroidal $B$-fields are confined to the crust
of the magnetar. If large enough ($B> B_{\rm cr}$), the magnetic field
will locally eliminate neutron superfluidity.  In particular,
according to Fig.~\ref{fig:Bcrit}, the neutron fluid in magnetars will
be non-superfluid (i.e., in a normal phase) for $B> 3\times 10^{17}$
G.  The non-superfluidity or partial superfluidity of magnetars will
clearly have profound implications for their neutrino emissivities,
transport properties, and thermal evolution, as well such dynamical
aspects as the damping of stellar oscillations and the interpretation
of rotational anomalies such as glitches and
anti-glitches.  {Note that the proton superconductivity
  in magnetar cores will be destroyed by Landau diamagnetic currents
  for fields even lower than those needed for the destruction of
  neutron $S$-wave
  superfluidity~\cite{2015PhRvC..91c5805S,2014arXiv1403.2829S}.  Of
  course the Pauli paramagnetic destruction mechanism discussed here
  for $S$-wave paired neutrons will apply to $S$-wave paired protons
  as well, but the diamagnetic mechanism is more important for
  protons.}

\section{Conclusions}
\label{sec:conclude}

 {
We have studied the phase diagram of dilute, spin-polarized neutron
matter with a BCS type order parameter.  Because two neutrons are unable
to form a bound pair in free space, there exists no {\it a priori} case for
the BEC of neutron-neutron pairs. However, the application of
Nozi{\`e}res-Schmitt-Rink theory~\cite{1985JLTP...59..195N} led us to
establishing a number of signatures in neutron matter that can be
interpreted as a precursor of BCS-BEC crossover; in the limit of zero
polarization these findings are in agreement with earlier studies of
this phenomenon~\cite{2006PhRvC..73d4309M,2007PhRvC..76f4316M,2009PhRvC..79e4003A,2013NuPhA.909....8S,2014PAN....77.1145K}.
}
Our conclusion can be summarized as follows:
\begin{enumerate}[(i)]
\item At low density, spin polarization does not affect the pairing
  substantially, but for higher densities and high polarizations, the
  pairing gap and hence the critical temperature $T_c$ are
  significantly suppressed. At finite polarization and low
  temperatures, we find a lower critical temperature that emerges from
  the combined effects of a polarization-induced separation and
  temperature-induced smearing of the Fermi surfaces
  involved.  {This feature tentatively indicates the
    possibility of the LOFF phase filling the low-temperature and
    high-density region of the phase diagram.}

\item We have analyzed some intrinsic features of the spin-polarized
  neutron condensate, specifically the gap, the kernel of the gap
  equation, the pair-condensate wave function, and the quasiparticle
  occupation numbers and energy spectra.  Similarities to
  behaviors found in a corresponding study of low-density
  isospin-asymmetric nuclear matter
  \cite{2012PhRvC..86f2801S,2014PhRvC..90f5804S} have been highlighted,
  along with their differences. We focus below on the principal findings
  of this analysis.

\item Under significant polarization, the kernel of the gap equation 
  acquires a double-peak structure in momentum space, in contrast 
  to the single peak present in the unpolarized case at the 
  Fermi momentum $k_F$. This feature is most pronounced in the high-density 
  and low-temperature limits.  Decreasing the density (or increasing 
  the temperature) smears out these structures. 

\item The Cooper-pair wave functions exhibit oscillatory behavior. At
  finite polarization the oscillations are in counterphase to those
  of the unpolarized case. The period of the oscillations is set
  by the wave vector as $2\pi/k_F$ and is not affected by the
  polarization. 

\item The quasiparticle occupation numbers show a separation of the 
  majority and minority spin populations by a breach around the 
  Fermi momentum $k_F$.  This is most pronounced in the high-density and 
  low-temperature limit, with the minority-spin component becoming 
  almost extinct. For high temperatures or low densities, this 
  breach is smeared out. 

\item Study of the quasiparticle dispersion relations establishes
  that they have a standard BCS form in the unpolarized case and 
  split into two branches at finite polarization, while retaining 
  the general BCS shape.  These spectra have minima at $k=k_F$, 
  as required. At large polarizations the energy spectrum of the 
  minority-spin particles crosses the zero-energy level, which is a 
  signature of gapless superconductivity. In other words, the 
  Fermi surface of the minority particles features locations 
  where modes can be excited without any energy cost.

\item At low densities, a relatively low magnetic 
  field is sufficient to generate a given polarization. In general, 
  the magnetic field required to produce a certain polarization 
  increases with decreasing temperature and with increasing 
  polarization. 

\item Finally, we have determined the critical field for unpairing of 
  the neutron condensate, which turns out to be in the range $B\sim 10^{17}$
  G. For larger fields the neutron fluid is non superfluid, which
  would have profound consequences for the thermal, rotational and
  oscillatory behavior of magnetars. 
\end{enumerate}

 Looking ahead, it should be mentioned that the
  present discussion does not take into account modifications of the
  pairing interaction in the medium, i.e., screening of the nuclear
  interaction.  In the case of unpolarized neutron matter, screening
  effects have been discussed extensively; see
  Ref.~\cite{2006pfsb.book..135S} and references cited therein, and
  especially Refs.~\cite{2007PhRvC..76f4316M,2010PhRvC..81c4007H} in
  the context of the BCS-BEC crossover. It is expected that pairing
  correlations are suppressed by the spin-fluctuation part of the
  screening interaction; hence the magnitude of the pairing gap and
  the range of densities over which pairing correlations extend will
  be reduced compared to what we find in the present study. This is
  strictly true if the spin-polarization does not change the sign of
  the screening interaction between neutrons. We anticipate that the
  changes will be of quantitative nature, without affecting the topology
  and the shape of the phase diagram of Fig.~\ref{phase_diagram_nn}.
  Accordingly, the main implication of the suppression of pairing by
  screening for the results we report is that the critical unpairing
  magnetic field obtained is an {\it upper bound}. A more complete
  application of our microscopic analysis to neutron star crusts would
  require the inclusion of nuclear clusters, as well as modifications of
  their properties induced by strong
  $B$ fields~\cite{2012PhRvC..86e5804C}. At those densities where
  apart from leading $S$-wave interaction, a subdominant $P$-wave
  interaction exists, the suppression of the $S$-wave pairing may give
  rise to $P$-wave superfluid, rather than normal spin-polarized
  fluid.  

Another relevant aspect of the many-body theory of this
  problem is that neutron matter is close to the unitary limit because
  of the large $nn$ scattering length.  Universal relations can be
  obtained in this limit, in particular for critical fields, with
  naive applications to neutron matter leading to
  estimates~\cite{2011PhRvB..83q4504H} consistent with those derived
  here.

\acknowledgments

M. S. acknowledges support from the HGS-HIRe graduate program at
Frankfurt University. A. S. is supported by the Deutsche
Forschungsgemeinschaft (Grant No. SE 1836/3-1) and by the NewCompStar
COST Action MP1304.  X. G. H. is supported by Fudan University Grant
EZH1512519 and Shanghai Natural Science Foundation Grant
No.14ZR1403000.  J. W. C.  acknowledges research support from the
McDonnell Center for the Space Sciences and expresses his thanks to
Professor Jos\'e Lu\'is da Silva and his colleagues at Centro de
Ci\^encias Matem\'aticas for gracious hospitality at the University of
Madeira.


\begin{thebibliography}{31}
\expandafter\ifx\csname natexlab\endcsname\relax\def\natexlab#1{#1}\fi
\expandafter\ifx\csname bibnamefont\endcsname\relax
  \def\bibnamefont#1{#1}\fi
\expandafter\ifx\csname bibfnamefont\endcsname\relax
  \def\bibfnamefont#1{#1}\fi
\expandafter\ifx\csname citenamefont\endcsname\relax
  \def\citenamefont#1{#1}\fi
\expandafter\ifx\csname url\endcsname\relax
  \def\url#1{\texttt{#1}}\fi
\expandafter\ifx\csname urlprefix\endcsname\relax\def\urlprefix{URL }\fi
\providecommand{\bibinfo}[2]{#2}
\providecommand{\eprint}[2][]{\url{#2}}

\bibitem[{\citenamefont{{Thompson} and {Duncan}}(1995)}]{1995MNRAS.275..255T}
\bibinfo{author}{\bibfnamefont{C.}~\bibnamefont{{Thompson}}} \bibnamefont{and}
  \bibinfo{author}{\bibfnamefont{R.~C.} \bibnamefont{{Duncan}}},
  \bibinfo{journal}{\mnras} \textbf{\bibinfo{volume}{275}},
  \bibinfo{pages}{255} (\bibinfo{year}{1995}).

\bibitem[{\citenamefont{{Nozi{\`e}res} and
  {Schmitt-Rink}}(1985)}]{1985JLTP...59..195N}
\bibinfo{author}{\bibfnamefont{P.}~\bibnamefont{{Nozi{\`e}res}}}
  \bibnamefont{and}
  \bibinfo{author}{\bibfnamefont{S.}~\bibnamefont{{Schmitt-Rink}}},
  \bibinfo{journal}{Journal of Low Temperature Physics}
  \textbf{\bibinfo{volume}{59}}, \bibinfo{pages}{195} (\bibinfo{year}{1985}).

\bibitem[{\citenamefont{{Matsuo}}(2006)}]{2006PhRvC..73d4309M}
\bibinfo{author}{\bibfnamefont{M.}~\bibnamefont{{Matsuo}}},
  \bibinfo{journal}{\prc} \textbf{\bibinfo{volume}{73}}, \bibinfo{eid}{044309}
  (\bibinfo{year}{2006}), \eprint{nucl-th/0512021}.

\bibitem[{\citenamefont{{Margueron} et~al.}(2007)\citenamefont{{Margueron},
  {Sagawa}, and {Hagino}}}]{2007PhRvC..76f4316M}
\bibinfo{author}{\bibfnamefont{J.}~\bibnamefont{{Margueron}}},
  \bibinfo{author}{\bibfnamefont{H.}~\bibnamefont{{Sagawa}}}, \bibnamefont{and}
  \bibinfo{author}{\bibfnamefont{K.}~\bibnamefont{{Hagino}}},
  \bibinfo{journal}{\prc} \textbf{\bibinfo{volume}{76}}, \bibinfo{eid}{064316}
  (\bibinfo{year}{2007}), \eprint{0710.4241}.

\bibitem[{\citenamefont{{Abe} and {Seki}}(2009)}]{2009PhRvC..79e4003A}
\bibinfo{author}{\bibfnamefont{T.}~\bibnamefont{{Abe}}} \bibnamefont{and}
  \bibinfo{author}{\bibfnamefont{R.}~\bibnamefont{{Seki}}},
  \bibinfo{journal}{\prc} \textbf{\bibinfo{volume}{79}}, \bibinfo{eid}{054003}
  (\bibinfo{year}{2009}), \eprint{0708.2524}.

\bibitem[{\citenamefont{{Sun} and {Pan}}(2013)}]{2013NuPhA.909....8S}
\bibinfo{author}{\bibfnamefont{B.~Y.} \bibnamefont{{Sun}}} \bibnamefont{and}
  \bibinfo{author}{\bibfnamefont{W.}~\bibnamefont{{Pan}}},
  \bibinfo{journal}{Nuclear Physics A} \textbf{\bibinfo{volume}{909}},
  \bibinfo{pages}{8} (\bibinfo{year}{2013}), \eprint{1304.6254}.

\bibitem[{\citenamefont{{Khodel} et~al.}(2014)\citenamefont{{Khodel}, {Clark},
  {Shaginyan}, and {Zverev}}}]{2014PAN....77.1145K}
\bibinfo{author}{\bibfnamefont{V.~A.} \bibnamefont{{Khodel}}},
  \bibinfo{author}{\bibfnamefont{J.~W.} \bibnamefont{{Clark}}},
  \bibinfo{author}{\bibfnamefont{V.~R.} \bibnamefont{{Shaginyan}}},
  \bibnamefont{and} \bibinfo{author}{\bibfnamefont{M.~V.}
  \bibnamefont{{Zverev}}}, \bibinfo{journal}{Physics of Atomic Nuclei}
  \textbf{\bibinfo{volume}{77}}, \bibinfo{pages}{1145} (\bibinfo{year}{2014}),
  \eprint{1310.5342}.

\bibitem[{\citenamefont{{Kanada-En'yo}
  et~al.}(2009)\citenamefont{{Kanada-En'yo}, {Hinohara}, {Suhara}, and
  {Schuck}}}]{2009PhRvC..79e4305K}
\bibinfo{author}{\bibfnamefont{Y.}~\bibnamefont{{Kanada-En'yo}}},
  \bibinfo{author}{\bibfnamefont{N.}~\bibnamefont{{Hinohara}}},
  \bibinfo{author}{\bibfnamefont{T.}~\bibnamefont{{Suhara}}}, \bibnamefont{and}
  \bibinfo{author}{\bibfnamefont{P.}~\bibnamefont{{Schuck}}},
  \bibinfo{journal}{\prc} \textbf{\bibinfo{volume}{79}}, \bibinfo{eid}{054305}
  (\bibinfo{year}{2009}), \eprint{0902.3717}.

\bibitem[{\citenamefont{{Alm} et~al.}(1993)\citenamefont{{Alm}, {Friman},
  {R{\"o}pke}, and {Schulz}}}]{1993NuPhA.551...45A}
\bibinfo{author}{\bibfnamefont{T.}~\bibnamefont{{Alm}}},
  \bibinfo{author}{\bibfnamefont{B.~L.} \bibnamefont{{Friman}}},
  \bibinfo{author}{\bibfnamefont{G.}~\bibnamefont{{R{\"o}pke}}},
  \bibnamefont{and} \bibinfo{author}{\bibfnamefont{H.}~\bibnamefont{{Schulz}}},
  \bibinfo{journal}{Nuclear Physics A} \textbf{\bibinfo{volume}{551}},
  \bibinfo{pages}{45} (\bibinfo{year}{1993}).

\bibitem[{\citenamefont{{Baldo} et~al.}(1995)\citenamefont{{Baldo}, {Lombardo},
  and {Schuck}}}]{1995PhRvC..52..975B}
\bibinfo{author}{\bibfnamefont{M.}~\bibnamefont{{Baldo}}},
  \bibinfo{author}{\bibfnamefont{U.}~\bibnamefont{{Lombardo}}},
  \bibnamefont{and} \bibinfo{author}{\bibfnamefont{P.}~\bibnamefont{{Schuck}}},
  \bibinfo{journal}{\prc} \textbf{\bibinfo{volume}{52}}, \bibinfo{pages}{975}
  (\bibinfo{year}{1995}).

\bibitem[{\citenamefont{{Sedrakian} and
  {Clark}}(2006{\natexlab{a}})}]{2006PhRvC..73c5803S}
\bibinfo{author}{\bibfnamefont{A.}~\bibnamefont{{Sedrakian}}} \bibnamefont{and}
  \bibinfo{author}{\bibfnamefont{J.~W.} \bibnamefont{{Clark}}},
  \bibinfo{journal}{\prc} \textbf{\bibinfo{volume}{73}}, \bibinfo{eid}{035803}
  (\bibinfo{year}{2006}{\natexlab{a}}), \eprint{nucl-th/0511076}.

\bibitem[{\citenamefont{{Huang}}(2010)}]{2010PhRvC..81c4007H}
\bibinfo{author}{\bibfnamefont{X.-G.} \bibnamefont{{Huang}}},
  \bibinfo{journal}{\prc} \textbf{\bibinfo{volume}{81}}, \bibinfo{eid}{034007}
  (\bibinfo{year}{2010}), \eprint{1002.0060}.

\bibitem[{\citenamefont{{Sun} et~al.}(2012)\citenamefont{{Sun}, {Sun}, and
  {Meng}}}]{2012PhRvC..86a4305S}
\bibinfo{author}{\bibfnamefont{T.~T.} \bibnamefont{{Sun}}},
  \bibinfo{author}{\bibfnamefont{B.~Y.} \bibnamefont{{Sun}}}, \bibnamefont{and}
  \bibinfo{author}{\bibfnamefont{J.}~\bibnamefont{{Meng}}},
  \bibinfo{journal}{\prc} \textbf{\bibinfo{volume}{86}}, \bibinfo{eid}{014305}
  (\bibinfo{year}{2012}), \eprint{1206.3407}.

\bibitem[{\citenamefont{{Lombardo} et~al.}(2001)\citenamefont{{Lombardo},
  {Nozi{\`e}res}, {Schuck}, {Schulze}, and {Sedrakian}}}]{2001PhRvC..64f4314L}
\bibinfo{author}{\bibfnamefont{U.}~\bibnamefont{{Lombardo}}},
  \bibinfo{author}{\bibfnamefont{P.}~\bibnamefont{{Nozi{\`e}res}}},
  \bibinfo{author}{\bibfnamefont{P.}~\bibnamefont{{Schuck}}},
  \bibinfo{author}{\bibfnamefont{H.-J.} \bibnamefont{{Schulze}}},
  \bibnamefont{and}
  \bibinfo{author}{\bibfnamefont{A.}~\bibnamefont{{Sedrakian}}},
  \bibinfo{journal}{\prc} \textbf{\bibinfo{volume}{64}},
  \bibinfo{pages}{064314} (\bibinfo{year}{2001}), \eprint{nucl-th/0109024}.

\bibitem[{\citenamefont{{Stein} et~al.}(2012)\citenamefont{{Stein}, {Huang},
  {Sedrakian}, and {Clark}}}]{2012PhRvC..86f2801S}
\bibinfo{author}{\bibfnamefont{M.}~\bibnamefont{{Stein}}},
  \bibinfo{author}{\bibfnamefont{X.-G.} \bibnamefont{{Huang}}},
  \bibinfo{author}{\bibfnamefont{A.}~\bibnamefont{{Sedrakian}}},
  \bibnamefont{and} \bibinfo{author}{\bibfnamefont{J.~W.}
  \bibnamefont{{Clark}}}, \bibinfo{journal}{\prc}
  \textbf{\bibinfo{volume}{86}}, \bibinfo{eid}{062801} (\bibinfo{year}{2012}),
  \eprint{1208.0123}.

\bibitem[{\citenamefont{{Stein} et~al.}(2014)\citenamefont{{Stein},
  {Sedrakian}, {Huang}, and {Clark}}}]{2014PhRvC..90f5804S}
\bibinfo{author}{\bibfnamefont{M.}~\bibnamefont{{Stein}}},
  \bibinfo{author}{\bibfnamefont{A.}~\bibnamefont{{Sedrakian}}},
  \bibinfo{author}{\bibfnamefont{X.-G.} \bibnamefont{{Huang}}},
  \bibnamefont{and} \bibinfo{author}{\bibfnamefont{J.~W.}
  \bibnamefont{{Clark}}}, \bibinfo{journal}{\prc}
  \textbf{\bibinfo{volume}{90}}, \bibinfo{eid}{065804} (\bibinfo{year}{2014}),
  \eprint{1410.1053}.

\bibitem[{\citenamefont{{Sedrakian} and
  {Clark}}(2006{\natexlab{b}})}]{2006pfsb.book..135S}
\bibinfo{author}{\bibfnamefont{A.}~\bibnamefont{{Sedrakian}}} \bibnamefont{and}
  \bibinfo{author}{\bibfnamefont{J.~W.} \bibnamefont{{Clark}}},
  \emph{\bibinfo{title}{{Nuclear Superconductivity in Compact Stars: BCS Theory
  and Beyond}}} (\bibinfo{publisher}{World Scientific Publishing Co},
  \bibinfo{year}{2006}{\natexlab{b}}), p. \bibinfo{pages}{135}.

\bibitem[{\citenamefont{{Ashcroft} and {Mermin}}(2005)}]{AM}
\bibinfo{author}{\bibfnamefont{N.~W.} \bibnamefont{{Ashcroft}}}
  \bibnamefont{and} \bibinfo{author}{\bibnamefont{{Mermin}}},
  \emph{\bibinfo{title}{{Introduction to Solid State Physics}}}
  (\bibinfo{publisher}{John Wiley \& Sons}, \bibinfo{year}{2005}).

\bibitem[{\citenamefont{{Sinha} and {Sedrakian}}(2015)}]{2015PhRvC..91c5805S}
\bibinfo{author}{\bibfnamefont{M.}~\bibnamefont{{Sinha}}} \bibnamefont{and}
  \bibinfo{author}{\bibfnamefont{A.}~\bibnamefont{{Sedrakian}}},
  \bibinfo{journal}{\prc} \textbf{\bibinfo{volume}{91}}, \bibinfo{eid}{035805}
  (\bibinfo{year}{2015}), \eprint{1502.02979}.

\bibitem[{\citenamefont{{Sinha} and {Sedrakian}}(2014)}]{2014arXiv1403.2829S}
\bibinfo{author}{\bibfnamefont{M.}~\bibnamefont{{Sinha}}} \bibnamefont{and}
  \bibinfo{author}{\bibfnamefont{A.}~\bibnamefont{{Sedrakian}}},
  \bibinfo{journal}{Physics of Particles and Nuclei}
  \textbf{\bibinfo{volume}{46}}, \bibinfo{eid}{1510} (\bibinfo{year}{2015}),
  \eprint{1403.2829}.

\bibitem[{\citenamefont{{Bertsch} and {Esbensen}}(1991)}]{1991AnPhy.209..327B}
\bibinfo{author}{\bibfnamefont{G.~F.} \bibnamefont{{Bertsch}}}
  \bibnamefont{and}
  \bibinfo{author}{\bibfnamefont{H.}~\bibnamefont{{Esbensen}}},
  \bibinfo{journal}{Annals of Physics} \textbf{\bibinfo{volume}{209}},
  \bibinfo{pages}{327} (\bibinfo{year}{1991}).

\bibitem[{\citenamefont{{Pastore} et~al.}(2013)\citenamefont{{Pastore},
  {Margueron}, {Schuck}, and {Vi{\~n}as}}}]{2013PhRvC..88c4314P}
\bibinfo{author}{\bibfnamefont{A.}~\bibnamefont{{Pastore}}},
  \bibinfo{author}{\bibfnamefont{J.}~\bibnamefont{{Margueron}}},
  \bibinfo{author}{\bibfnamefont{P.}~\bibnamefont{{Schuck}}}, \bibnamefont{and}
  \bibinfo{author}{\bibfnamefont{X.}~\bibnamefont{{Vi{\~n}as}}},
  \bibinfo{journal}{\prc} \textbf{\bibinfo{volume}{88}}, \bibinfo{eid}{034314}
  (\bibinfo{year}{2013}), \eprint{1303.5651}.

\bibitem[{\citenamefont{{Zhang} et~al.}(2014)\citenamefont{{Zhang}, {Matsuo},
  and {Meng}}}]{2014PhRvC..90c4313Z}
\bibinfo{author}{\bibfnamefont{Y.}~\bibnamefont{{Zhang}}},
  \bibinfo{author}{\bibfnamefont{M.}~\bibnamefont{{Matsuo}}}, \bibnamefont{and}
  \bibinfo{author}{\bibfnamefont{J.}~\bibnamefont{{Meng}}},
  \bibinfo{journal}{\prc} \textbf{\bibinfo{volume}{90}}, \bibinfo{eid}{034313}
  (\bibinfo{year}{2014}), \eprint{1310.3625}.

\bibitem[{\citenamefont{{Hagino} et~al.}(2010)\citenamefont{{Hagino}, {Sagawa},
  and {Schuck}}}]{2010JPhG...37f4040H}
\bibinfo{author}{\bibfnamefont{K.}~\bibnamefont{{Hagino}}},
  \bibinfo{author}{\bibfnamefont{H.}~\bibnamefont{{Sagawa}}}, \bibnamefont{and}
  \bibinfo{author}{\bibfnamefont{P.}~\bibnamefont{{Schuck}}},
  \bibinfo{journal}{Journal of Physics G Nuclear Physics}
  \textbf{\bibinfo{volume}{37}}, \bibinfo{eid}{064040} (\bibinfo{year}{2010}),
  \eprint{0912.4792}.

\bibitem[{\citenamefont{{Su} et~al.}(1987)\citenamefont{{Su}, {Yang}, and
  {Kuo}}}]{1987PhRvC..35.1539S}
\bibinfo{author}{\bibfnamefont{R.~K.} \bibnamefont{{Su}}},
  \bibinfo{author}{\bibfnamefont{S.~D.} \bibnamefont{{Yang}}},
  \bibnamefont{and} \bibinfo{author}{\bibfnamefont{T.~T.~S.}
  \bibnamefont{{Kuo}}}, \bibinfo{journal}{\prc} \textbf{\bibinfo{volume}{35}},
  \bibinfo{pages}{1539} (\bibinfo{year}{1987}).

\bibitem[{\citenamefont{{Chabanat} et~al.}(1998)\citenamefont{{Chabanat},
  {Bonche}, {Haensel}, {Meyer}, and {Schaeffer}}}]{1998NuPhA.635..231C}
\bibinfo{author}{\bibfnamefont{E.}~\bibnamefont{{Chabanat}}},
  \bibinfo{author}{\bibfnamefont{P.}~\bibnamefont{{Bonche}}},
  \bibinfo{author}{\bibfnamefont{P.}~\bibnamefont{{Haensel}}},
  \bibinfo{author}{\bibfnamefont{J.}~\bibnamefont{{Meyer}}}, \bibnamefont{and}
  \bibinfo{author}{\bibfnamefont{R.}~\bibnamefont{{Schaeffer}}},
  \bibinfo{journal}{Nuclear Physics A} \textbf{\bibinfo{volume}{635}},
  \bibinfo{pages}{231} (\bibinfo{year}{1998}).

\bibitem[{\citenamefont{Haidenbauer and Plessas}(1984)}]{PhysRevC.30.1822}
\bibinfo{author}{\bibfnamefont{J.}~\bibnamefont{Haidenbauer}} \bibnamefont{and}
  \bibinfo{author}{\bibfnamefont{W.}~\bibnamefont{Plessas}},
  \bibinfo{journal}{Phys. Rev. C} \textbf{\bibinfo{volume}{30}},
  \bibinfo{pages}{1822} (\bibinfo{year}{1984}).

\bibitem[{\citenamefont{{He} et~al.}(2006)\citenamefont{{He}, {Jin}, and
  {Zhuang}}}]{2006PhRvB..74u4516H}
\bibinfo{author}{\bibfnamefont{L.}~\bibnamefont{{He}}},
  \bibinfo{author}{\bibfnamefont{M.}~\bibnamefont{{Jin}}}, \bibnamefont{and}
  \bibinfo{author}{\bibfnamefont{P.}~\bibnamefont{{Zhuang}}},
  \bibinfo{journal}{\prb} \textbf{\bibinfo{volume}{74}}, \bibinfo{eid}{214516}
  (\bibinfo{year}{2006}), \eprint{cond-mat/0606322}.

\bibitem[{\citenamefont{{Gubankova} et~al.}(2003)\citenamefont{{Gubankova},
  {Liu}, and {Wilczek}}}]{2003PhRvL..91c2001G}
\bibinfo{author}{\bibfnamefont{E.}~\bibnamefont{{Gubankova}}},
  \bibinfo{author}{\bibfnamefont{W.~V.} \bibnamefont{{Liu}}}, \bibnamefont{and}
  \bibinfo{author}{\bibfnamefont{F.}~\bibnamefont{{Wilczek}}},
  \bibinfo{journal}{Physical Review Letters} \textbf{\bibinfo{volume}{91}},
  \bibinfo{eid}{032001} (\bibinfo{year}{2003}), \eprint{hep-ph/0304016}.

\bibitem[{\citenamefont{{Chamel} et~al.}(2012)\citenamefont{{Chamel}, {Pavlov},
  {Mihailov}, {Velchev}, {Stoyanov}, {Mutafchieva}, {Ivanovich}, {Pearson}, and
  {Goriely}}}]{2012PhRvC..86e5804C}
\bibinfo{author}{\bibfnamefont{N.}~\bibnamefont{{Chamel}}},
  \bibinfo{author}{\bibfnamefont{R.~L.} \bibnamefont{{Pavlov}}},
  \bibinfo{author}{\bibfnamefont{L.~M.} \bibnamefont{{Mihailov}}},
  \bibinfo{author}{\bibfnamefont{C.~J.} \bibnamefont{{Velchev}}},
  \bibinfo{author}{\bibfnamefont{Z.~K.} \bibnamefont{{Stoyanov}}},
  \bibinfo{author}{\bibfnamefont{Y.~D.} \bibnamefont{{Mutafchieva}}},
  \bibinfo{author}{\bibfnamefont{M.~D.} \bibnamefont{{Ivanovich}}},
  \bibinfo{author}{\bibfnamefont{J.~M.} \bibnamefont{{Pearson}}},
  \bibnamefont{and}
  \bibinfo{author}{\bibfnamefont{S.}~\bibnamefont{{Goriely}}},
  \bibinfo{journal}{\prc} \textbf{\bibinfo{volume}{86}}, \bibinfo{eid}{055804}
  (\bibinfo{year}{2012}), \eprint{1210.5874}.

\bibitem[{\citenamefont{{He} and {Zhuang}}(2011)}]{2011PhRvB..83q4504H}
\bibinfo{author}{\bibfnamefont{L.}~\bibnamefont{{He}}} \bibnamefont{and}
  \bibinfo{author}{\bibfnamefont{P.}~\bibnamefont{{Zhuang}}},
  \bibinfo{journal}{\prb} \textbf{\bibinfo{volume}{83}}, \bibinfo{eid}{174504}
  (\bibinfo{year}{2011}), \eprint{1101.5694}.

\end{thebibliography}
\end{document}